\documentclass[12pt]{iopart}
\usepackage{epsfig}

\begin{document}
\title{Counterrotating perfect fluid discs as sources of electrovacuum
static spacetimes}

\author{Gonzalo Garc\'{\i}a-Reyes.\footnote{e-mail: ggr1970@yahoo.com} and Guillermo
A. Gonz\'{a}lez\footnote{e-mail: guillego@uis.edu.co}}

\address{Escuela de F\'{\i}sica, Universidad Industrial de Santander, A. A.
678, Bucaramanga, Colombia}

\begin{abstract}
The interpretation of some electrovacuum spacetimes in terms  of
counterrotating perfect fluid discs is presented. The  interpretation is mades
by means of an ``inverse problem''  approach used to obtain disc sources of
known static solutions of the Einstein-Maxwell equations. In order to do such
interpretation, a detailed study is presented  of the  counterrotating model
(CRM) for generic electrovacuum static  axially symmetric rela\-tivistic thin
discs with nonzero  radial pressure. Four simple families of models of 
counterrotating charged discs based on Chazy-Curzon-type,  Zipoy-Voorhees-type,
Bonnor-Sackfield-type, and charged and magnetized Darmois  electrovacuum
metrics are considered  where we obtain some discs with a CRM well behaved.
\end{abstract}

\submitto{CQG}
\pacs{04.20.-q, 04.20.Jb, 04.40.Nr}

\section{Introduction}

Several methods are known to exactly solve the Einstein and  Einstein-Maxwell
field equations, or to generate new exact  solutions from simple known
solutions \cite{KSHM}. However, the above mentioned methods in general lead to 
solutions without a clear physical interpretation or to  solutions that depend
of many parameters without a clear physical meaning. Accordingly, it is of
importance to have  some appropriate procedures to obtain physical
interpretations of these exact solutions. So, in the past years such  
procedures have been developed for static and stationary  axially symmetric
solutions in terms of thin and, more  recently, thick disc models.

Stationary or static axially symmetric exact solutions of Einstein  equations
describing relativistic thin discs are of great astrophysical importance since
they can be used as models of certain stars, galaxies and accretion discs.
These were first studied by Bonnor and Sackfield \cite{BS}, obtaining
pressureless static discs, and then by Morgan and Morgan, obtaining static
discs with and without radial pressure \cite{MM1,MM2}. In connection with
gravitational collapse, discs were first studied by Chamorro, Gregory, and
Stewart \cite{CHGS}. Discs with radial tension have been also studied
\cite{GL1}.

Several classes of exact solutions of the Einstein field equations
corresponding to static and stationary thin discs have been obtained by
different authors, with or without radial pressure (see refs. \cite{LP} to
\cite{GL2}). Except for the pressureless disks all the other disks have as
source matter with azimuthal pressure (tension)  different from the radial
pressure (tension).  However, in some cases these disks can be interpreted as
the superposition of two counterrotating perfect fluids.  A detailed study of
the counterrotating model for the case of static thin disks  is  presented in
\cite{GE}. Recently, more realistic models of thin disks and thin disks with
halos  made of perfect fluids were considered in \cite{VL1}. 

Disc sources for stationary axially symmetric spacetimes with magnetic fields
are also of astrophysical importance mainly in the study of neutron stars,
white dwarfs and galaxy formation. Although discs with electric fields do not
have clear astrophysical importance, their study may be of interest in the
context of exact solutions. Thin discs  have been discussed as sources for
Kerr-Newman fields \cite{LBZ}, magnetostatic axisymmetric fields \cite{LET1},
conformastationary metrics \cite{KBL}, and electrovacuum static counterrotating
discs \cite{GG}. Following  Ref. \cite{LBZ} the resulting discs can  be
interpreted either as rings with internal pressure and currents or as two
counterrotating  streams of freely moving charged particles, i.e., which move
along electrogeodesics (solution to the geodesic equation in the presence of a
Lorentz force).

In all the above cases, the discs are obtained by an ``inverse problem''
approach, called by Synge the ``{\it g-method}'' \cite{SYN}. The method works
as follows: a solution of the vacuum Einstein equations is taken, such that
there is a discontinuity in the derivatives of the metric tensor on the plane
of the disc, and the energy-momentum tensor is obtained from the Einstein 
equations. The physical properties of the matter distribution are then studied
by an analysis of the surface energy-momentum tensor so obtained. Very recently
the above procedure has been generalized in order to obtain thick disc models
\cite{GL3} and also non-axisymmetric planar distributions of charged dust
\cite{VL2}. As we can see, by this ``inverse problem'' approach we can do
physical interpretations of many known solutions of the  Einstein and
Einstein-Maxwell vacuum equations, in the sense that the thin disks can act as
exact sources for the space-time metrics given by the vacuum solutions.

The purpose  of this paper is the interpretation of some static electrovacuum
spacetimes in terms of counterrotating perfect fluid discs. In order to do such
interpretation, a detailed study is presented  of the counterrotating model
(CRM) for generic electrovacuum static axially symmetric rela\-tivistic thin
discs with nonzero radial pressure. The counterrotating model for the case of
static thin discs with radial pressure was recently studied in \cite{GE} and 
for static discs without radial pressure but without electromagnetic field in
\cite{GG}, so the material presented here is a continuation of the mentioned
works. The  paper is structured as follows. In Sec. \ref{sec:discs},  we
present a summary of the procedure to construct   models of  discs  with
nonzero radial pressure. We obtain expressions for the surface energy-momentum
tensor and the surface current density of the disc.
 
In Sec. \ref{sec:crm}, the discs are interpreted in terms of the CRM. We find a
general constraint over the counterrotating tangential velocities needed to
cast the surface energy-momentum tensor of the disc as the superposition of two
counterrotating  charged perfect fluids. We also find expressions for the
energy density, pressure,  current density, and tangential velocity of the
counterrotating fluids. We show that, in general, it is not possible to take
the two counterrotating fluids as circulating along electrogeodesics nor take
the two counterrotating tangential velocities as equal and opposite.

In Sec. \ref{sec:examp}, four simple families of models of  counterrotating
charged discs based on Chazy-Curzon-type,  Zipoy-Voorhees-type,
Bonnor-Sackfield-type, and charged and magnetized Darmois metrics are presented
where we obtain some discs with a well behaved CRM. In particular, we study
the  tangential velocities, energy and electric charge densities in the
particular case when both streams have  equal pressure or  equal energy
density. Also the stability against radial  perturbation is analyzed. Finally,
in Sec. \ref{sec:discu},  we summarize our main results.   

\section{Electrovacuum static relativistic discs}\label{sec:discs}

\subsection{Models of discs with nonzero radial pressure}

In this section we present a summary of the procedure to obtain electrovacuum 
static axially symmetric relativistic thin discs with nonzero radial pressure.
The  line element of  a static axisymmetric field can be written in
quasi-cylindrical coordinates (t,$\varphi,r,z$) in the form  
\begin{equation}
ds^2 = - \ e^{2 \Psi} dt^2 + e^{- 2 \Psi} [R^2 d\varphi^2 + e^{2 \Lambda} (dr^2
+ dz^2)] \  , \label{eq:met}
\end{equation}
where $\Psi$, $\Lambda$ and $R$ are functions  of $r$ and $z$ only. The vacuum
Einstein-Maxwell equations, in  geometrized units in which $8 \pi G = c = \mu
_0 = \varepsilon _0 = 1$,  are given by 
\begin{eqnarray}
&   &    R_{ab} \  =  \ T_{ab}, \label{eq:einmax1} \\
&   &    F^{ab}_{ \ \ \ ; b} = 0,  \label{eq:einmax2}
\end{eqnarray}
with
\begin{eqnarray}
&   &     T_{ab} \  =  \ F_{ac}F_b^{ \ c} - \frac 14 g_{ab}F_{cd}F^{cd},
\label{eq:et}  \\
&   &   F_{ab} =  \partial _ a A_b -  \partial _ b A_a.\label{eq:fab}
\end{eqnarray}

In order to obtain a solution of (\ref{eq:einmax1}) -- (\ref{eq:einmax2})
representing a thin disc at $z=0$, we assume that the components of the metric
tensor are continuous across the disc, but with first derivatives discontinuous
on the plane $z=0$, with discontinuity functions
$$
b_{ab} \ = g_{ab,z}|_{_{z = 0^+}} \ - \ g_{ab,z}|_{_{z = 0^-}} \ = \ 2 \
g_{ab,z}|_{_{z = 0^+}} \ .
$$
Thus, the Einstein-Maxwell equations yield an energy-momentum tensor $T_{ab}  =
T^{\mathrm {elm}}_{ab} + T^{\mathrm {mat}}_{ab}$, where $T^{\mathrm {mat}}_{ab}
= Q_{ab} \ \delta (z)$, and a  current density $J_a = j_a  \delta (z) = - 2
e^{2 (\Psi - \Lambda)} A_{a,z} \delta (z)$,  where  $\delta (z)$ is the usual
Dirac function with support on the disk. $T^{\mathrm {elm}}_{ab}$ is the
electromagnetic tensor defined in Eq.  (\ref{eq:et}), $j_a$ is the current
density on the plane  $z=0$, and
$$
Q^a_b = \frac{1}{2}\{b^{az}\delta^z_b - b^{zz}\delta^a_b + g^{az}b^z_b -
g^{zz}b^a_b + b^c_c (g^{zz}\delta^a_b - g^{az}\delta^z_b)\}
$$
is the distributional energy-momentum tensor.

The ``true'' surface energy-momentum tensor (SEMT) of the disc, $S_{ab}$, and
the ``true'' surface current density, $\mbox{\sl j}_a$, can be obtained through
the relations
\begin{eqnarray}
&&S_{ab} \ = \ \int T^{\mathrm {mat}}_{ab} \ ds_n \ = \ e^{\Lambda - \Psi} \
Q_{ab} \ , \\
&&\mbox{\sl j}_a \ = \ \int J_a \ ds_n \ = \ e^{\Lambda - \Psi} \ j_a  \ ,
\end{eqnarray}
where $ds_n = \sqrt{g_{zz}} \ dz$ is the ``physical measure'' of length in the
direction normal to the disc. For the metric (\ref{eq:met}), the  nonzero
components of  $S_a^b$  are
\begin{eqnarray}
&S^0_0 &= \ 2 e^{\Psi - \Lambda} \left[ \Lambda,_z - \ 2 \Psi,_z + \
\frac{R,_z}{R} \right ] , \label{eq:emt1}     		\\
&S^1_1 &= \ 2 e^{\Psi - \Lambda} \Lambda,_z , \label{eq:emt2} \\
&S^2_2 &= \ 2 e^{\Psi - \Lambda} \left [ \frac{R,_z}R \right] , \label{eq:emt3}
\label{eq:emt}\end{eqnarray}
and the components of the surface current density are equal to
\begin{eqnarray}
& \mbox{\sl j}_t &= \ -2 e^{\Psi - \Lambda} \psi _{,z} , \label{eq:corelec}  \\
&\mbox{\sl j}_{\varphi} &= \ -2 e^{\Psi - \Lambda} A _{,z},  \label{eq:cormag}
\label{eq:cor}\end{eqnarray}
where $\psi$ and $A$ are the electrostatic and  magnetostatic potentials,
respectively, which  are also functions  of $r$ and $z$ only. All the
quantities are evalua\-ted at $z = 0^+$.

In order to give physical significance  to the components of  the
energy-momentum tensor  $S_a^b$ we project it onto the   orthonormal tetrad 
${{\rm e}_{\hat a}}^b = \{ V^b , W^b , X^b,Y^b \}$, where
\begin{eqnarray}
V^a = e^{- \Psi}( 1, 0, 0, 0 ), & \qquad & X^a = e^{\Psi - \Lambda}( 0, 0, 1, 0
), \nonumber \\
&& \label{eq:tetrad} \\
W^a = \frac{e^\Psi}R( 0, 1, 0, 0 ), & \qquad & Y^a = e^{\Psi - \Lambda}( 0, 0,
0, 1 ). \nonumber
\end{eqnarray}
In terms of this tetrad (or observer ) the metric, the SEMT and the current
density  may be decomposed as  
\begin{eqnarray}
g_{ab} &=& - V_a V_b + W_a W_b + X_a X_b + Y_a Y_b  , \label{eq:metdia} \\
S_{ab} &=& \ \epsilon V_a V_b + p_\varphi W_a W_b + p_r X_a X_b ,
\label{eq:emtdia} \\
\mbox{\sl j}_a &=& \ \sigma V_a + \mbox{\sl j} W_a, \label{eq:ja}
\end{eqnarray}
where
\begin{equation}
\epsilon \ = \ - S^0_0, \quad  p_\varphi \ = \ S^1_1, \quad  p_r \ = \ S^2_2, 
\end{equation}
are, respectively, the energy density, the azimuthal pressure, the radial
pressure,
\begin{equation}
\sigma = \mbox{\sl j}^0/V^0, \quad \mbox{\sl j}  = \mbox{\sl j}^1/W^1, 
\label{eq:dps}
\end{equation}
are the electric charge density and the azimutal current density  of the disc
measured by this observer with four-velocity $V^a$, and
\begin{equation}
\varrho \ = \ \epsilon \ + \ p_\varphi \ + \ p_r 
\end{equation}
is the ``effective Newtonian density''.

\subsection{Solution of the field equations}

For the metric (\ref{eq:met}), the  Einstein-Maxwell equations in vacuum imply
that $R$ satisfies  the Laplace's  equation
\begin{equation}
R_{,rr} \ + \ R_{,zz} \ = \ 0,  \label{eq:lap}
\end{equation}
so that the function $R$ can be considered as the real part of an analytical  
function  $F (\nu) = R(r,z) + i Z(r,z)$, where $\nu = r + iz$. Thus the
function  $F (\nu)$ defines  a conformal transformation
\begin{equation}
r \rightarrow  R (r, z) , \qquad z \rightarrow Z (r, z), \label{eq:conf}
\end{equation}
in such a way that the metric (\ref{eq:met}) takes the Weyl form
\begin{equation}
ds^2 = - \ e^{2 \mu} dt^2 + e^{- 2 \mu} [R^2 d\varphi^2 + e^{2 \lambda} (dR^2 +
dZ^2)],  \label{eq:newmet}
\end{equation}
where $R$ and $Z$ are the Weyl's canonical coordinates.

In these coordinates,  the field equations (\ref{eq:einmax1}) --
(\ref{eq:einmax2}) are equivalent to the usual complex  Ernst equations
\cite{E2}
\begin{eqnarray}
f \Delta {\cal E} &=& (\nabla {\cal E} + 2 \Phi^\ast \nabla \Phi) \cdot
\nabla{\cal E},  \label{eq:ernst1}    \\
f \Delta \Phi &=& (\nabla {\cal E} + 2 \Phi^\ast \nabla \Phi) \cdot \nabla\Phi,
\label{eq:ernst2}
\end{eqnarray}
where  $\Delta$ and  $\nabla$ are  the  standard differential operators in
cylindrical coordinates, $f= e^{2\mu}$ and ${\cal E}={\cal E}^*$ (static
spacetime). The metric functions are obtained via 
\begin{eqnarray}
f  &=& {\cal E} +\Phi \Phi^* ,   \\
\lambda _{,\zeta}&=& \sqrt 2 \frac R f \left[ \frac {1}{4f} ( {\cal E}_{,\zeta}
+2 \Phi ^* \Phi_{,\zeta} )({\cal E}_{,\zeta} +2 \Phi  \Phi^*_{,\zeta} ) -
\Phi_{,\zeta}\Phi^*_{,\zeta}\right], \label{eq:lambda}
\end{eqnarray}
where $\sqrt 2\zeta = R +iZ$, so that $\sqrt 2 \partial_{,\zeta}=
\partial_{,R}-i\partial_{,Z}$. The electromagnetic potentials are related to 
$\Phi$ via
\begin{equation} 
A_{\varphi,\zeta} =  \sqrt 2 i \frac R f ( {\rm Im} \Phi )_{,\zeta},
\end{equation}  
and $A_t = \sqrt 2{\rm Re} \Phi$.

It is known that in static electromagnetic fields there is a  lineal relation
between the electrostatic potential and its  magnetic counterpart
\cite{Per,Das}. Some authors therefore study only electric fields. In this
paper we consider the field equations when both electric and magnetic fields
are presents, since the true magnetic potential will be  different, and
therefore the physical properties will differ too. 
 
Once a solution of the system of equations (\ref{eq:ernst1}) --
(\ref{eq:ernst2}) is known, we can obtain a solution of the field equations
(\ref{eq:einmax1}) -- (\ref{eq:einmax2}) in the original coordinates  by
setting 
\begin{eqnarray}
R(r,z) &=&   {\rm Re} F(\nu), \label{eq:comf1}       \\
\Psi (r,z) \ &=& \ \mu (R,Z) , \label{eq:com3}\\
\Lambda (r,z) \ &=& \ \lambda (R,Z) \ + \ \ln |F' (\nu)| , \label{eq:com4} \\
\psi (r,z) \ &=& \ A_t (R,Z),  \\
A (r,z) \ &=& \ A_\varphi (R,Z), \label{eq:comf6}
\end{eqnarray}%
where  $F'=dF/d\nu$. With this transformation $p_r$ can be written as 
\begin{equation}
p_r = \frac {2 e^{\mu - \lambda}}{\sqrt {R_{,r}^2 +R_{,z}^2 }} \frac
{R_{,z}}{R}.
\end{equation}

Thus, the construction of thin discs models with nonzero radial pressure
depends strongly of the analytical function $F (\nu)$, due to the fact that
this function gives the behavior of the radial pressure and also defines the
conformal transformation  that leads the field equations to the  form given for
Ernst  (\ref{eq:ernst1}) -- (\ref{eq:ernst2}). A common  choice for $F (\nu)$
is
\begin{equation}
F ( \nu ) \ = \ \nu \ + \ i z_0 \ , \label{eq:dcr}
\end{equation}
where $z_0$ is a constant. In this case we have infinite discs with zero radial
pressure. This is the well known ``displace, cut and reflect'' method used in
almost all of the works about discs models to generate thin discs from
solutions of the Einstein and Einstein-Maxwell equations. Another possible
choice for $F (\nu)$ was presented in reference \cite{GL1} and is given by
\begin{equation}
F (\nu) \ = \ \nu + \alpha \sqrt{\nu^2 - 1} \ , \label{eq:fnu}
\end{equation}
where $\alpha \geq 0$. With this choice  for $F(\nu)$ the general expresion for
the radial pressure leads to
\begin{equation}
p_r = \frac {2 \alpha e^{\mu - \lambda} }{[1 + ( \alpha ^2 -1)r^2]^{1/2} } ,
\end{equation}
so that we obtain thin discs with nonzero radial pressure and of finite radius,
located at $z = 0$, $0 \leq r \leq 1$. In order to obtain discs with non-unit
radius, we only need to make the transformation $r \rightarrow a r$, where $a$
is the radius of the disc. We shall also take $F(\nu)$ as given by
(\ref{eq:fnu}).

With this choice for $F(\nu)$ the image of the disc by the  conformal mapping
(\ref{eq:conf}) is the surface 
\begin{equation}
\alpha ^2 R^2 + Z^2 = \alpha ^2
\end{equation}
so that the discs are mapped into spheroidal thin shells of  matter and its
exterior  is mapped into the exterior of the  shells. We have three possible
values for $\alpha$: $\alpha =  1$, a spherical shell, $\alpha > 1$, a prolate
spheroidal  shell, and  $0 < \alpha < 1$, an oblate spheroidal shell.  By
considering the above shells as sources, we then seek for exterior solutions of
the  field equations (\ref{eq:ernst1}) -- (\ref{eq:ernst2}) adapted to the
symmetry of the shells.  Then, using (\ref{eq:comf1}) -- (\ref{eq:comf6}) we 
obtain the corresponding disc solutions in the original  coordinates.  

\section{The counterrotating model}\label{sec:crm}

\subsection{Counterrotating charged perfect fluid discs}

We now consider, based on Refs. \cite{LET2} and \cite{FMP}, the possibility
that the SEMT $S^{ab}$ and the current density  $\mbox{\sl j}^a$ can be written
as the superposition of two counterrotating charged perfect fluids that
circulate in opposite directions; that is, we assume 
\begin{eqnarray}
S^{ab} &=& S_+^{ab} \ + \ S_-^{ab} \ , \label{eq:emtsum}   \\
\mbox{\sl j}^a    &=& \mbox{\sl j}_+^a + \mbox{\sl j}_-^a, \label {eq:corsum} 
\end{eqnarray}
where the  quantities on the right-hand side  are, respectively, the SEMT and
the current density of the prograde and retrograde counterrotating fluids. 

Let  $U_\pm^a = ( U_\pm^0 , U_\pm^1, 0 , 0 )$ be the velocity vectors of the
two counterrotating fluids. In order to do the decomposition (\ref{eq:emtsum})
and (\ref{eq:corsum}) we project the velocity vectors onto the tetrad ${{\rm
e}_{\hat a}}^b$, using the relations \cite{CHAN}
\begin{equation}
U_\pm^{\hat a} \ = \ {{\rm e}^{\hat a}}_b U_\pm^b, \quad  \quad U_\pm^ a = \ 
{{\rm e}_{\hat b}}^a U_\pm^{\hat b} .
\end{equation}
In terms of  the tetrad (\ref{eq:tetrad}) we can write
\begin{equation}
U_\pm^a \ = \ \frac{ V^a + v_\pm W^a }{\sqrt{1 - v_\pm^2}} , \label{eq:vels}
\end{equation}
where $v_\pm = U_\pm^{\hat 1} / U_\pm^{\hat 0}$ are the tangential velocities
of the streams with respect to the tetrad.

Another quantity related with the counterrotating motion is the specific
angular momentum of a particle rotating at a radius $r$, defined as $h_\pm =
g_{\varphi\varphi} U_\pm^\varphi$. We can write
\begin{equation}
h_\pm \ = \ \frac{R e^{- \Psi} v_\pm}{\sqrt{1 - v_\pm^2}}. \label{eq:moman}
\end{equation}
This quantity can be used to analyze the stability of the discs against radial
perturbations. The condition of stability,
\begin{equation}
\frac{d(h^2)}{dr} \ > \ 0  ,
\end{equation}
is an extension of Rayleigh criteria of stability of a fluid in rest in a
gravitational field \cite{FLU}.

In terms of the metric $h_{ab} = g_{ab} - Y_a Y_b$ of the hypersurface $z = 0$,
we can write the SEMT as
\begin{equation}
S^{ab} \ = \ ( \epsilon + p_r ) V^a V^b \ + \ ( p_\varphi - p_r ) W^a W^b \ + \
p_r h^{ab} ,
\end{equation}
and  using (\ref{eq:vels}) we obtain
\begin{eqnarray}
S^{ab} & = & \frac{ f( v_- , v_- ) (1 - v_+^2) \ U_+^a U_+^b }{(v_+ - v_-)^2}
\nonumber	\\
&	&		\nonumber	\\
& + & \frac{ f( v_+ , v_+ ) (1 - v_-^2) \ U_-^a U_-^b
}{(v_+ - v_-)^2}			\nonumber	\\
&	&		\nonumber	\\
& - & \frac{ f( v_+ , v_- ) (1 - v_+^2)^{\frac{1}{2}} (1 - v_-^2)^{\frac{1}{2}}
( U_+^a U_-^b + U_-^a U_+^b ) }{(v_+ - v_-)^2}		\nonumber	\\
&	&		\nonumber	\\
& + & p_r h^{ab} ,	\nonumber
\end{eqnarray}
where
\begin{equation}
f( v_1 , v_2 ) \ = \ ( \epsilon + p_r ) v_1 v_2 + p_\varphi - p_r.
\label{eq:fuu}
\end{equation}
Clearly, in order to cast the SEMT in the form (\ref{eq:emtsum}), the mixed
term must be absent and therefore the counterrotating tangential velocities
must be related by
\begin{equation}
f( v_+ , v_- ) \ = \ 0  , \label{eq:liga}
\end{equation}
where we assume that $|v_\pm| \neq 1$.

Then, assuming a given choice for the tangential velocities in agreement with
the above relation, we can write the SEMT as  (\ref{eq:emtsum}) with
\begin{equation}
S^{ab}_\pm = ( \epsilon_\pm + p_\pm ) \ U_\pm^a U_\pm^b \ + \ p_\pm \ h^{ab} ,
\end{equation}
so that we have two counterrotating perfect fluids with energy densities and
pressures, measured by an observer comoving with the streams, given by 
\begin{eqnarray}
\epsilon_\pm + \ p_\pm &= & \left[ \frac{ 1 - v_\pm^2 }{v_\mp - v_\pm} \right] 
( \epsilon + p_r ) v_\mp  , \label{eq:epcon1} \\
p_+ + \ p_- &= &  \ p_r . \label{eq:epcon2}
\end{eqnarray}
From these equations also follows that 
\begin{equation}
\epsilon_+ + \ \epsilon_- = \ \ \epsilon \ + \ p_r \ - \ p_\varphi.
\end{equation}
Thus the SEMT $S^{ab}$ can be written as the superposition of two
counterrotating perfect fluids if, and only if, the constraint (\ref{eq:liga}) 
admits a solution such that $v_+ \neq v_-$. This result is completely
equivalent to the necessary and sufficient condition obtained in Ref.
\cite{FMP}.

Similarly, substituting  (\ref{eq:vels}) in (\ref{eq:ja}) we can write the
current density as (\ref{eq:corsum}) with
\begin{equation}
\mbox{\sl j}^a_\pm  = \sigma _\pm U_\pm^a, 
\end{equation}
where $\sigma _\pm$ are the counterrotating electric charge densities  measured
by an observer comoving with the streams,   which are given by
\begin{equation}
\sigma _{ \pm} =  \left[ \frac { \sqrt{1-v^2_\pm }} {  v_\pm -v_\mp} \right] (
\mbox{\sl j} - \sigma v_\mp  ). \label{eq:sig} 
\end{equation}
Thus, we have a disc  makes of two counterrotating charged perfect fluids with
energy densities and pressures given by (\ref{eq:epcon1}) and
(\ref{eq:epcon2}), and electric charge densities given by (\ref{eq:sig}).

\subsection{Counterrotating tangential velocities}

All the main quantities associated with the CRM depend on the counterrotating
tangential velocities $v_\pm$. However, even for known values of $v_\pm$, the
system of equations (\ref{eq:epcon1}) -- (\ref{eq:epcon2}) is  underdetermined
so that we can have information about the physical quantities corresponding 
to  each  stream only in some particular situations. We will examine two simple
cases in this paper, when the two streams have equal pressure  and when they
have equal energy density. On the other hand, the constraint (\ref{eq:liga})
does not determine $v_\pm$ uniquely so that we need to imposse some additional
requeriment in order to obtain a complete determination of the tangential
velocities leading  to a well defined CRM.

A possibility, commonly assumed, is to take the two counterrotating streams as
circulating along electrogeodesics
\begin{equation}
\frac 12 \epsilon _\pm g_{ab,r}U^a_\pm U^b_\pm = - \sigma _\pm F_{ra} U^a_\pm.
\label{eq:elecgeo}
\end{equation}
Let $\omega_\pm = U_\pm^1/U_\pm^0$ be the angular velocities of the particles.
In terms of $\omega_\pm$  (\ref{eq:elecgeo}) takes the form 
\begin{equation}
\frac 12 \epsilon _\pm ( g_{11,r} \omega _\pm ^2 + g_{00,r} ) = - \sigma _{\pm}
\frac 1 {U_\pm ^0} ( \psi _{,r} + A_{,r} \omega _{\pm}), \label{eq:elecgeo1}
\end{equation}
and $v_\pm$ can be written as
\begin{equation}
v_\pm \ =  \  \left[ \frac{V^0}{W^1} \right] \omega_\pm.
\end{equation}

As an example, let us consider the electrogeodesic motion of the particles 
when we have equal energy density. By simplicity, we will analyze the 
electrostatic and magnetostatic cases  separately.  The angular velocity
$\omega_\pm$ can be obtained from the electrogeodesic equation
(\ref{eq:elecgeo1}) and, in the electrostatic case, we obtain $\omega _\pm =
\pm \omega$ with
\begin{equation}
\omega ^2 = \frac {e^{4 \Psi}}R \left [ \frac { -2\Psi_{,r}(\Psi_{,z}-\Lambda
_{,z}) + \psi _{,r} \psi _{,z}e^{-2 \Psi} } { -2(\Psi _{,z}-\Lambda
_{,z})(R_{,r} -R \Psi_{,r}) + R \psi _{,r} \psi _{,z}e^{-2 \Psi} } \right ], 
\end{equation}
and therefore
\begin{equation}
f( v_+ , v_- ) = -R \left [ \frac { 2(\Psi_{,z}-\Lambda _{,z}) dp_r /dr + p_r
\psi _{,r} \psi _{,z} e^{-2 \Psi}  } { -2(\Psi_{,z}-\Lambda _{,z})(R_{,r} -R
\Psi_{,r}) + R \psi _{,r} \psi _{,z}e^{-2 \Psi} } \right ], 
\end{equation} 
where we have used  the Einstein-Maxwell equations (\ref{eq:einmax1}) and
(\ref{eq:einmax2}) for $\Lambda$  and the expressions (\ref{eq:emt1}) -
(\ref{eq:cormag}) for the SEMT and the  current density. It follows immediately
that $f( v_+ , v_- )$ vanishes  if
\begin{equation} 
\frac 1 {p_r} \frac {dp_r}{dr}  =  - \frac {\psi _{,r} \psi _{,z}e^{-2
\Psi}}{2(\Psi_{,z}-\Lambda _{,z})},
\end{equation}
and therefore
\begin{equation}
p_r = c_1 e^{f_1(r)}, \label{eq:conprel}
\end{equation}
where $c_1$ is a constant  and 
\begin{equation}
f_1(r) = - \int \frac { \psi_{,r} \psi_{,z} e^{- 2 \Psi } }{ 2(\Psi_{,z}-
\Lambda_{,z}) } dr. 
\end{equation}

Similarly, for magnetostatic fields  we obtain again $\omega _\pm = \pm \omega$
with
\begin{equation}
\omega ^2 = e^{4 \Psi} \left [ \frac { 2\Psi_{,r}(\Psi_{,z}-\Lambda _{,z}) + A
_{,r} A _{,z}e^{2 \Psi} / R^2 } { 2R(\Psi_{,z}-\Lambda _{,z})(R_{,r} -R
\Psi_{,r}) +  A _{,r} A _{,z} e ^{2 \Psi} } \right ], 
\end{equation}
so that
\begin{equation}
f( v_+ , v_- ) = R \left [ \frac { 2(\Psi_{,z}-\Lambda _{,z}) dp_r/dr - p_r A
_{,r}  A _{,z} e^{2 \Psi}  } { -2(\Psi_{,z}-\Lambda _{,z})(R_{,r} -R \Psi_{,r})
+ A _{,r} A _{,z}e^{2 \Psi}/R } \right ]. 
\end{equation}  
That is, $f( v_+ , v_- )$ vanishes if
\begin {equation}
\frac 1 {p_r} \frac {dp_r}{dr}  =   \frac {A _{,r} A _{,z}e^{2 \Psi}}{2(
\Psi_{,z}-\Lambda _{,z})},
\end{equation}
and therefore
\begin{equation}
p_r = c_2 e^{f_2(r)}, \label{eq:conprmag} 
\end{equation}
where again $c_2$ is a constant  and 
\begin{equation}
f_2 (r)= \int \frac { A_{,r} A_{,z} e^{ 2 \Psi } }{ 2(\Psi_{,z} - \Lambda_{,z})
} d r. 
\end{equation}
That is, for  electrostatic or magnetostatic fields and equal  energy densities
the streams circulate along electrogeodesics  only if the radial pressure
satisfies (\ref{eq:conprel}) or  (\ref{eq:conprmag}), respectively. Note that
the two  electrogeodesic fluids are circulating with equal and opposite
tangential velocities. In  more general situations, for  example with
simultaneous presence of electric and magnetic  fields, $p_r$ is  a 
complicated function of $r$ that can be  obtained  by a similar procedure and
the streams would move  with different velocities. 

As we can see, in order to have the counterrotating fluids circulating along
electrogeodesics, the radial pressure must satisfy a very strong condition
that, in general, is not satisfied by the thin disc solutions. Accordingly, we
need to consider an additional requeriment different of the electrogeodesic
motion in order to exactly determine the counterrotating velocities. Another
possibility is to take the particles of the streams not circulating along
electrogeodesics but with equal and opposite tangential velocities, that is
\begin{equation}
v_\pm \ = \ \pm \ v ,
\end{equation}
so that (\ref{eq:liga}) is equivalent to
\begin{equation}
v^2 \ = \ \left[ \frac{p_\varphi - p_r}{\epsilon + p_r} \right] . \label{eq:u2}
\end{equation}
This choice, commonly considered, leads  to a complete determination of the
velocity vectors. However, this can be made only when the above expression is
positive definite. In particular, when the two streams have equal pressure or
equal energy density we have $p_\pm=\frac{1}{2} p_r$ and   $\epsilon_\pm=\frac
12 (\epsilon + p_r - p_\varphi)$.

In the general case, when (\ref{eq:u2}) is not positive definite, the two
fluids circulate with different velocities and we can write (\ref{eq:liga}) as
\begin{equation}
v_+ v_- \ = \ \left[ \frac{p_r - p_\varphi}{\epsilon + p_r} \right],
\end{equation}
and so we can obtain a CRM only if
\begin{equation}
| \ p_r - p_\varphi \ | \ < \ | \ \epsilon + p_r \ | . \label{eq:vabs}
\end{equation}
However, this relation does not determine completely the tangential velocities,
and therefore the CRM is undetermined. 
 
\section{Some simple examples of counterrotating charged 
discs}\label{sec:examp}

\subsection{Discs from a Chazy-Curzon-type solution}

The first family of solutions to consider is a  Chazy-Curzon-type solution
which is given by
\begin{eqnarray}
e^\mu    &=& \frac {2}{(a+1) e^{\gamma/\rho} -  (a-1) e^{-\gamma/\rho }} ,
\label{eq:ccl1} \\
\lambda &=&  - \frac{\gamma^2 \sin^2 \theta}{2\rho^2} , \\
A_t    &=&  \frac{\sqrt{2} q_1 [e^{\gamma/\rho} - e^{-\gamma/\rho}]}{(a+1)
e^{\gamma/\rho} - (a-1) e^{-\gamma/\rho}} , \\
A_\varphi & = &\sqrt 2 \gamma q_2 \cos \theta,
\end{eqnarray}
where $a^2=1+b^2$, with  $b^2 = q_1^2 + q_2^2 $, and $\gamma$  is a real
constant. Here $q_1$ and $q_2$ are the  electric and magnetic parameters,
respectively. $\rho$ and $\theta$ are  quasi-spherical coordinates, related to
the Weyl coordinates  by 
\begin{equation}
R   = \rho \sin \theta ,  \quad  \quad Z = \rho \cos \theta,
\end{equation}
where $0 \leq \rho \leq \infty$ and $- \frac \pi 2 \leq \theta \leq \frac \pi
2$. In this coordinates  we must choose  $\alpha = 1$ and the shell would be
located at $\rho = 1$.  This solution can be generated, in these coordinates, 
using  the well-known  complex potential formalism proposed by Ernst  \cite{E2}
from the   Chazy-Curzon vacuum solution \cite{CH,C}, by choosing the parameter
$q$ of Ref. \cite{ E2} as complex. For  $q_2=0$ we have an electrostatic
solution and for $q_1=0$ one  obtains its  magnetostatic analogue \cite{B1}.

From the above expressions we can compute the physical  quantities associated
with the discs. We obtain 
\begin{eqnarray}
\epsilon &=& \varrho -  p_r -  p_ \varphi,   \\
p_\varphi &= & p_r (1+\gamma ^2 r^2),      \\
p_r &=&  p_0 e^{\gamma ^2 r^2 /2} , \\
\sigma & = & -    \frac{q_1 p_0}{\sqrt 2}   \ p_r,
\end{eqnarray} 
where
\begin{equation}
p_0 = \frac {4}{(a+1)e^\gamma - (a-1)e^{-\gamma}}
\end{equation}
is a  constant  and
\begin{equation}
\varrho =  \frac{\gamma p_0}{2}    [(a+1)e^\gamma + (a-1)e^{-\gamma} ]  p_r.
\label{eq:rho}
\end{equation}
The strong energy condition requires that $\varrho \geq 0 $, so that we must
choose $\gamma > 0$. We can see that $p_0$ is a positive constant for these
values of $\gamma$.

\begin{figure}
$$
\begin{array}{cc}
\epsilon  &  p_\varphi, \ \ p_r \\
\epsfig{width=2.2in,file=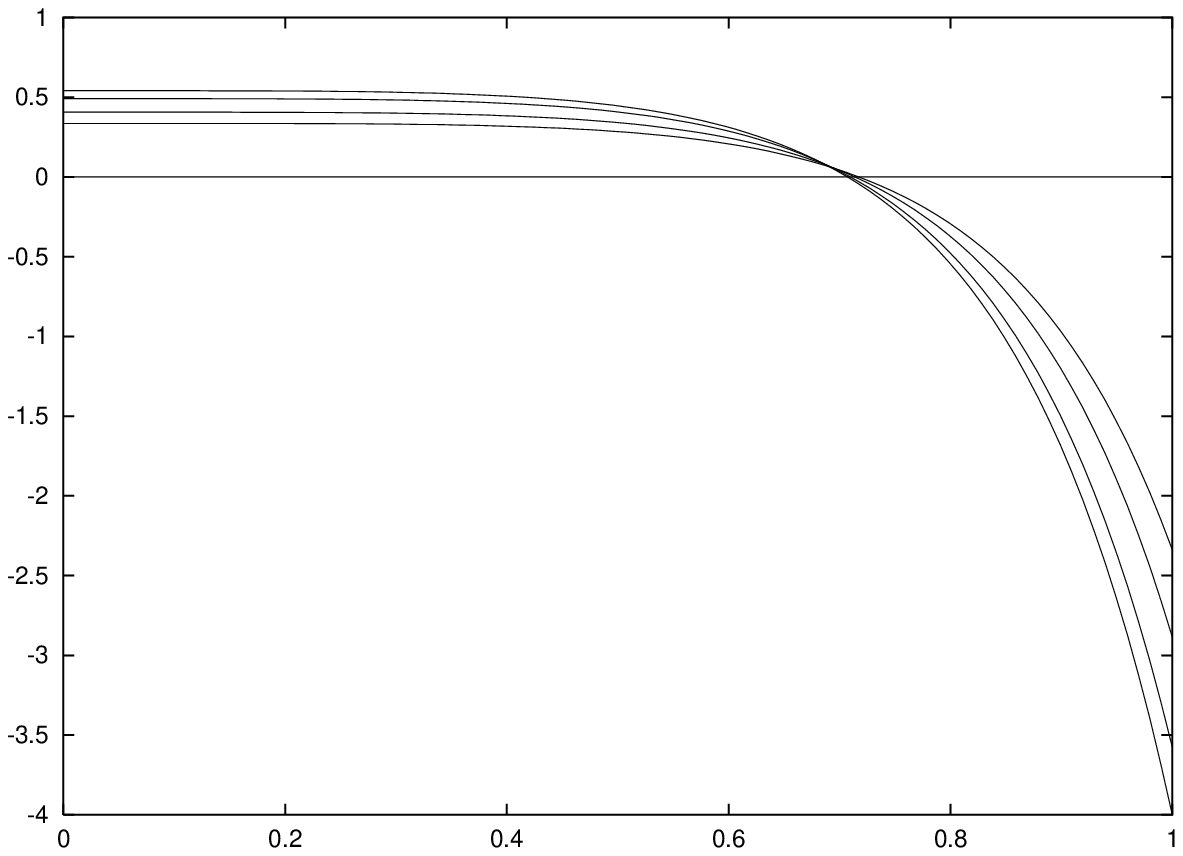} & \epsfig{width=2.2in,file=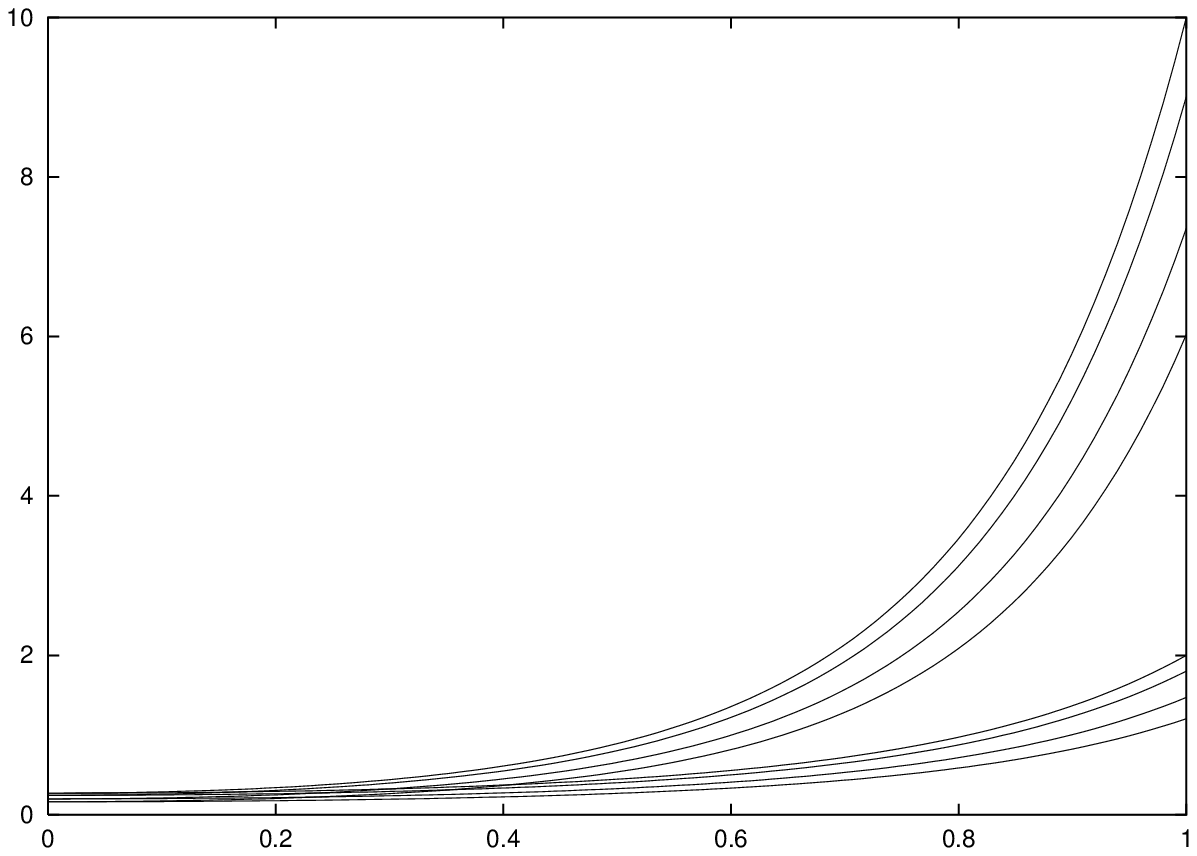} \\
(a)     &   (b)
\end{array}
$$	
$$
\begin{array}{cc}
-\sigma  &  v^2 \\
\epsfig{width=2.2in,file=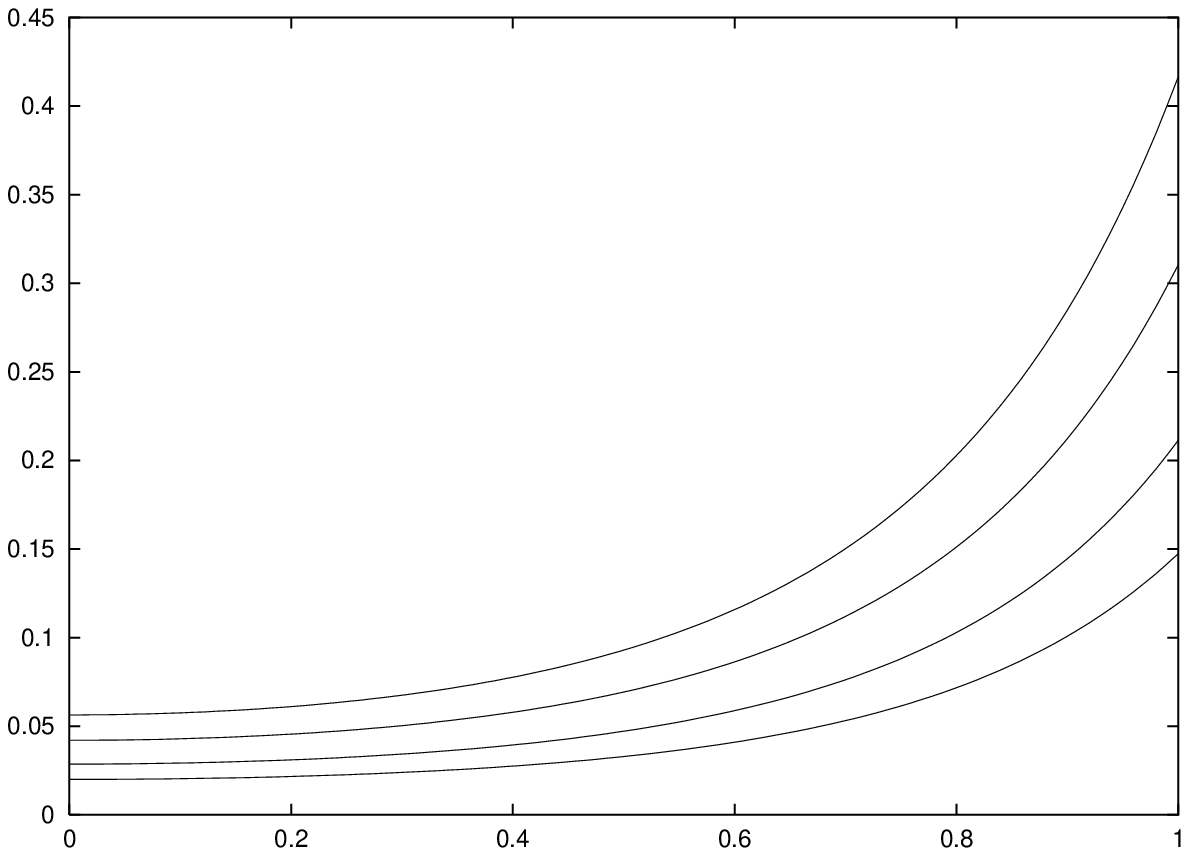} & \epsfig{width=2.2in,file=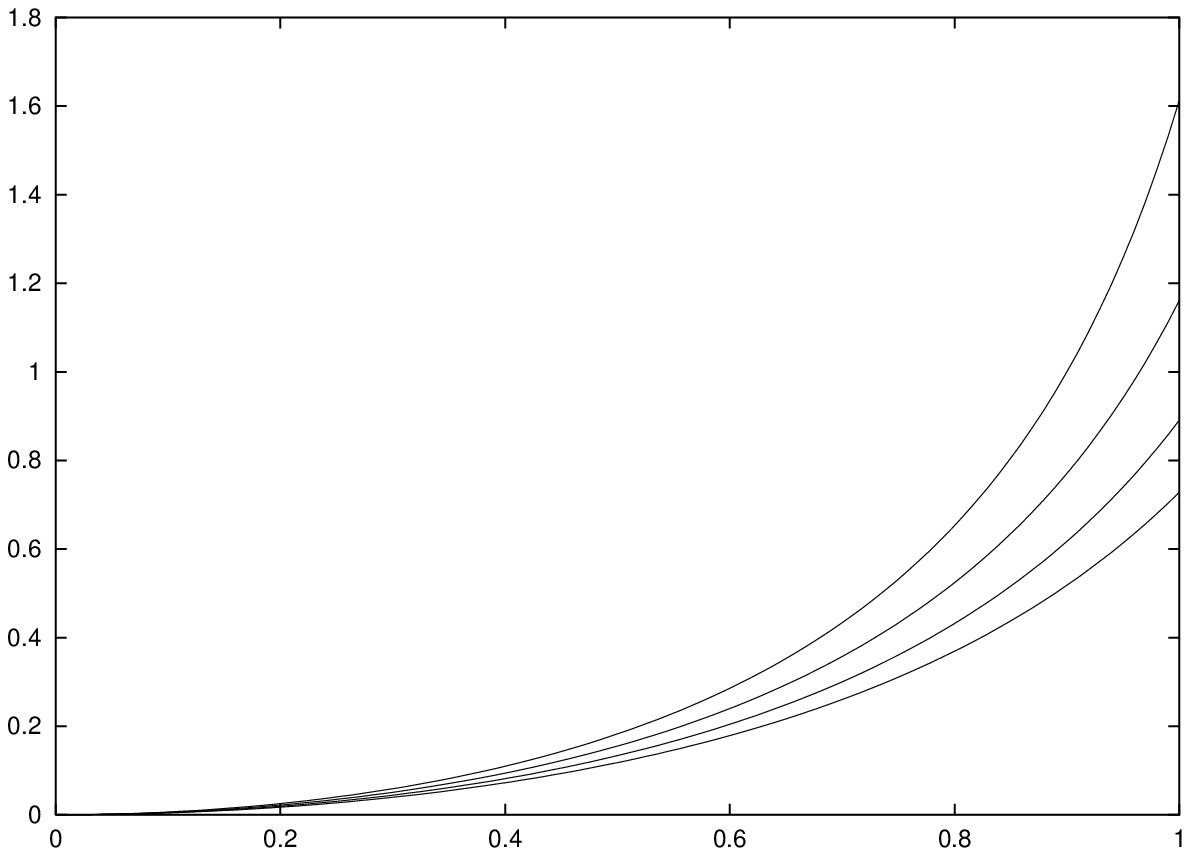} \\
(c)     &   (d) 
\end{array}
$$	
$$
\begin{array}{cc}
 h^2 &  \epsilon _\pm, \ \ -\sigma _\pm \\
\epsfig{width=2.2in,file=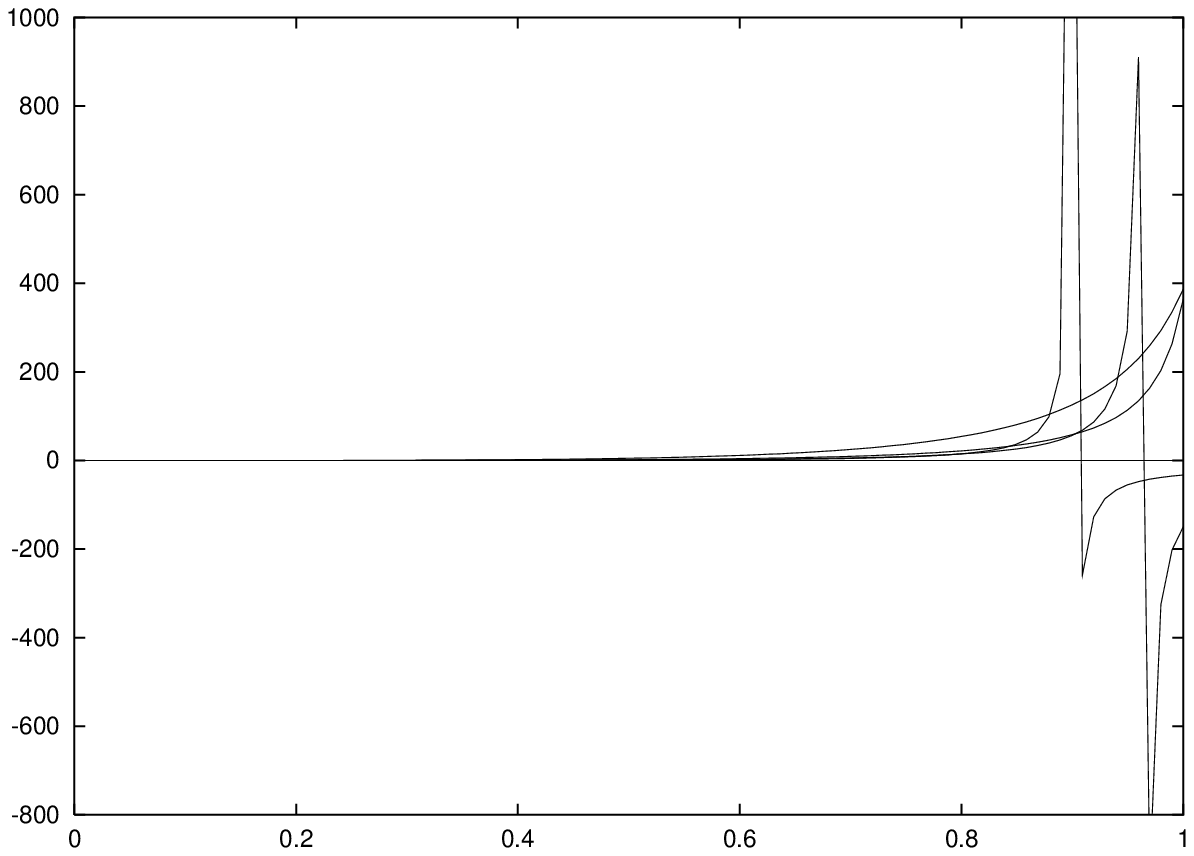} & \epsfig{width=2.2in,file=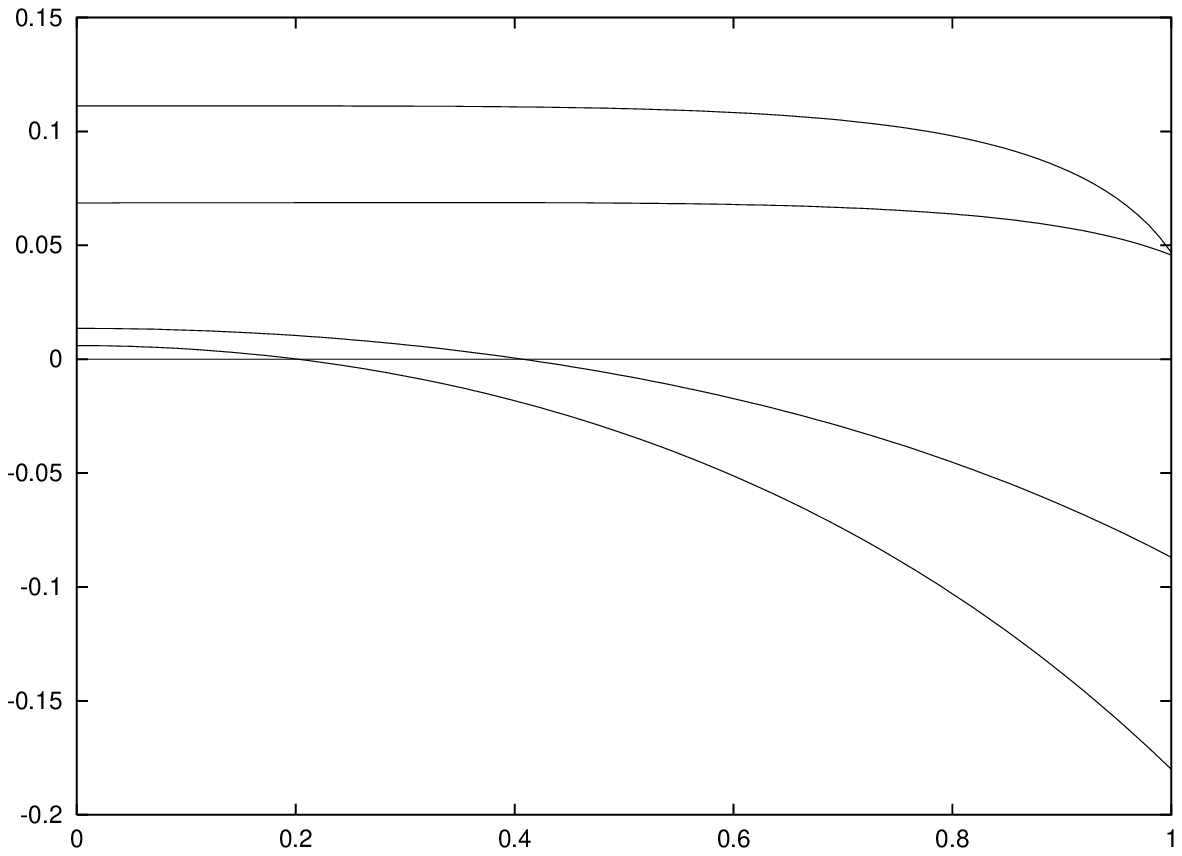}\\
(e)     &   (f)
\end{array}
$$	
\caption{For Chazy-Curzon-type fields we plot, as functions of  $r$, $(a)$ the
energy density  $\epsilon$ and  $(b)$ the  pressures $p_\varphi$ (upper curves)
and $p_r$ for discs  with $\gamma=2$ and $q_1=q_2=0$ (top curves), $0.5$,
$1.0$, and  $1.5$ (bottom curves), $(c)$ the electric charge  density  
$\sigma$ for discs with $\gamma=2$ and $q_1=q_2=0$ (axis r),  $0.2$, $0.3$,
$0.5$, and $1.5$ (top curve), $(d)$ $v^2$ for   discs  with $\gamma=0.7$ and 
$q_1=q_2=2$ (top curve), $3$,   $5$, and $10$ (bottom curve), $(e)$ the angular
momentum   $h^2$  for discs  with $\gamma=0.7$ and  $q_1=q_2=2$ , $3$  (sharp
curves), $5$, and $10$ (top curve), and $(f)$  $\epsilon _\pm$ (lower curves)
and $\sigma _\pm$ for discs   with $\gamma=0.7$ and  $q_1=q_2=5$ and $10$ (top
and bottom  curves, respectively).} \label{fig:cc}
\end{figure}

In order to study the behavior of these quantities we perform a graphical
analysis of them  for discs  with $\gamma=2$ and different values of $q_1$ and
$q_2$, as functions of $r$. In Figs. \ref{fig:cc}$(a)$ and \ref{fig:cc}$(b)$ we
show the energy density and the pressures $p_r$ and $p_\varphi$ for  $q_1 = q_2
= 0$, $0.5$, $1.0$, and $1.5$. We see that the energy density  initially is a
positive quantity  and then becomes negative in violation of the weak energy
condition.  Therefore  only the central region of these discs has a physically
reazonable behavior.   We can observe that the pressures $p_r$ and $p_\varphi$ 
are always positive quantities everywhere on discs. We also see  that  the
presence of  electromagnetic  field diminishes  the energy density  near to the
center of the discs  and  later increases it, and it diminishes the pressures
everywhere on discs.   Next, the charge density $\sigma$  is  represented in
Fig. \ref{fig:cc}$(c)$ for  $q_1=q_2=0$, $0.2$, $0.3$, $0.5$, and $1.5$. As we
can see, it exhibits a similar behavior to the radial pressure.  We also
computed these functions  for other values of the parameters (with $\gamma >0$)
and, in all the cases, we found a similar behavior.

We now consider the  CRM. All the significant quantities can also be expressed
in analytic form from the above expressions but the results are so cumbersome
that better turns out only to analyze them graphically. Let us   consider the
case when the two fluids move with equal and opposite tangential velocities. 
In Fig. \ref{fig:cc}$(d)$ we plot the tangential velocity curves of the
counterrotating streams $v^2$  for discs  with $\gamma=0.7$ and $q_1=q_2=2$,
$3$,  $5$, and $10$.  We  see that the inclusion of  electromagnetic field can
make the velocities of the particles smaller than the light velocity. In Fig.
\ref{fig:cc}$(e)$  we have drawn the  angular momentum  $ h^2$  for the same
values of the parameters.  In some cases we have strong changes in the  slope
at certain value of $r$, which means that there is a strong instability there.
For other values of the parameters $ h^2$ is an increasing monotonous function
of $r$ that corresponds to a stable CRM for the discs. Finally, in Fig.
\ref{fig:cc}$(f)$ we have plotted the energy and electric charge densities
$\epsilon _\pm$ and  $\sigma _\pm$ in the particular case   when the two
streams have equal pressure or equal energy density, for values of the
parameters for which $v^2<1$. We see that $\sigma _\pm$ behaves in the opposite
way to  $\sigma$, whereas  $\epsilon _\pm$ has a similar behavior to the energy
density $\epsilon$. Therefore, for  Chazy-Curzon-type fields we can build 
counterrotating thin disc sources only with a well-behaved central region.

\subsection{Discs from a Zipoy-Voorhees-type solution}

The second  family of  solutions considered  is a Zipoy-Voorhees-type solution
which can be written as 
\begin{eqnarray}
e^\mu    &=& \frac{2 (x^2 - 1)^{\gamma/2}}{(a +1)(x + 1)^\gamma - (a-1)(x -
1)^\gamma} , \label{eq:zv1} \\
\lambda &=&\frac{\gamma ^2}{2} \ln \left [ \frac {x^2-1}{x^2-y^2}  \right ], \\
A_t &=& \frac { \sqrt 2 q_1 [(x + 1)^\gamma - (x - 1)^\gamma] }{(a+1)(x +
1)^\gamma -(a-1)(x - 1)^\gamma}, \\
A_\varphi & = &\sqrt 2 k\gamma q_2 y, \label{eq:zv4}
\end{eqnarray}
where $a^2=1+b^2$, with  $b^2 = q_1^2 + q_2^2 $, and $\gamma$ is a real
constant. Here $q_1$ and $q_2$ are again the  electric and magnetic parameters,
respectively. $x$ and $y$ are the prolate spheroidal coordinates, related to
the Weyl coordinates by 
\begin{equation}
R^2   = k^2 (x^2-1)(1-y^2),  \quad  \quad Z  = kxy, \label{eq:coorp}
\end{equation}
where  $1 \leq x \leq \infty$ and $0  \leq y \leq 1 $. In this case we must
choose  $\alpha >1$ and the shell would be located at $x =  \alpha / k > 1$ and
$y= \sqrt {1-r^2}$, with  $k= \sqrt {\alpha ^2 -1}$.

This solution can also be generated, in these coordinates,  using the
well-known  complex potencial formalism proposed by Ernst \cite{E2} from the  
Zipoy-Voorhees vacuum solution \cite{Z,V}, also known as the Weyl
$\gamma$-solution \cite{W1,W2}, by choosing the parameter $q$ of Ref. \cite{
E2} as complex.  For  $q_2=0$ we also have an electrostatic solution and for
$q_1=0$  one  obtains its  magnetostatic equivalent \cite{B1}. The case $q_2=0$
and $\gamma =1$ corresponds to the   Reissner-Nordstr\"{o}m  solution
\cite{KSHM}, in which case $a= m/k$, $q_1=e/k$, with $k^2=m^2-e^2$, so that
$a^2=1+q_1^2$,  $m$ and $e$ being the mass and  charge parameters,
respectively. When    $q_1=0$ and $\gamma =2$ one arrives to  a Taub-NUT-type 
magnetostatic solution.   This   solution can also   be generated, in these
coordinates,  using a well-known theorem  proposed by Bonnor (see Ref.
\cite{B2}) from  the Taub-NUT  vacuum solution. Note that for $b=0$, it reduces
to the Darmois metric \cite{KSHM}.

From the above expressions we can compute the physical quantities associated
with the discs. We obtain
\begin{eqnarray}
\epsilon &=& \varrho - p_r -p_ \varphi,   \\
p_ \varphi &=& \left[ \frac {1+ k^2 \gamma^2 r^2}{1+ k^2 r^2}  \right] p_r, \\
p_r &=& p_0 [1+ k^2 r^2]^{\frac 12 (\gamma^2-1)}, \\
\sigma &=&  -  \frac {  \gamma k q_1  p_0}{\sqrt 2 \alpha ^2} \ p_r, 
\end{eqnarray}
where 
\begin{equation}
p_0 = \frac { 4 \alpha}{ (a+1)(\alpha + k)^\gamma - (a-1)(\alpha - k)^\gamma }
\end{equation}
is a constant and 
\begin{equation}
\varrho = \frac { \gamma k p_0}{2 \alpha ^2} \left[(a+1)(\alpha + k)^\gamma +
(a-1)(\alpha - k)^\gamma \right ]  p_r.
\end{equation}
Again   we must choose $\gamma >0$ so that $\varrho \geq 0$  and the solution
satisfies the strong energy condition. $p_0$ is also  a positive constant for
these values of $\gamma$. 

\begin{figure}
$$
\begin{array}{cc}
\epsilon  &  p_\varphi, \ \ p_r \\
\epsfig{width=2.2in,file=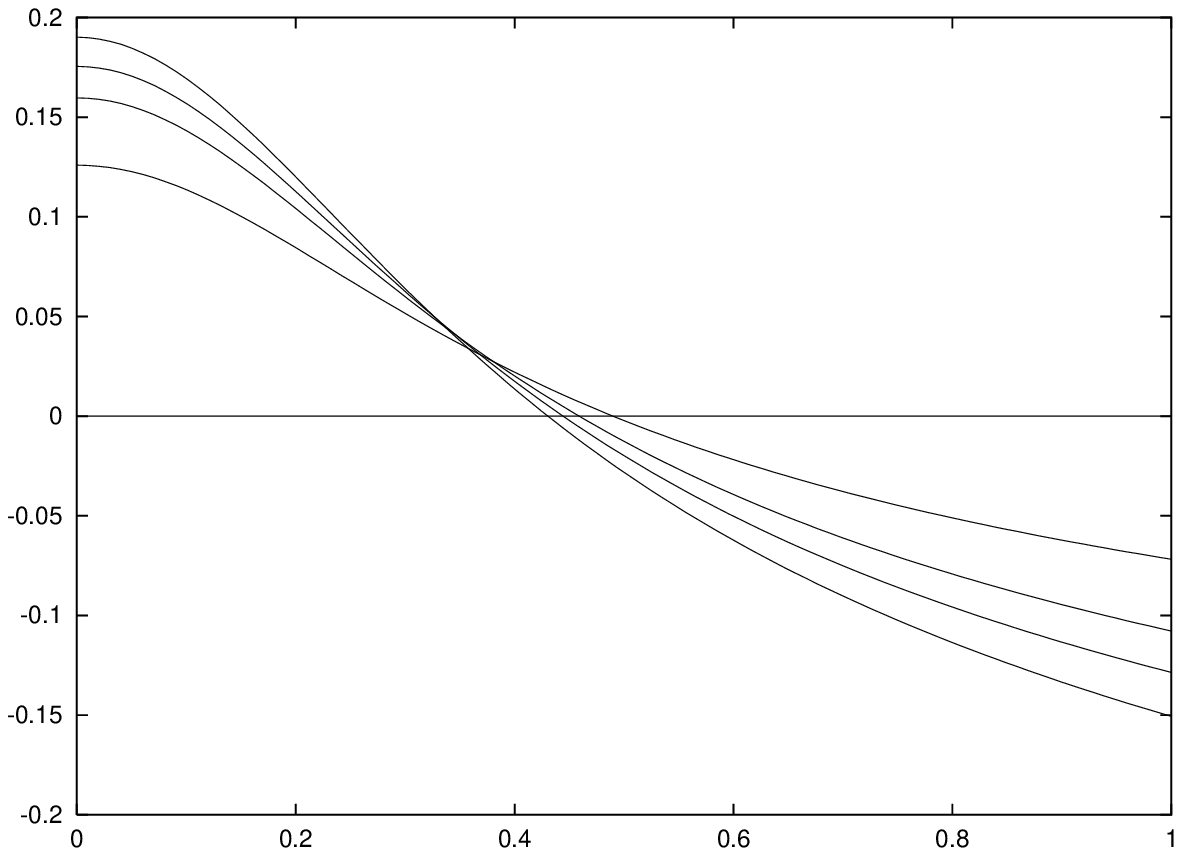} & \epsfig{width=2.2in,file=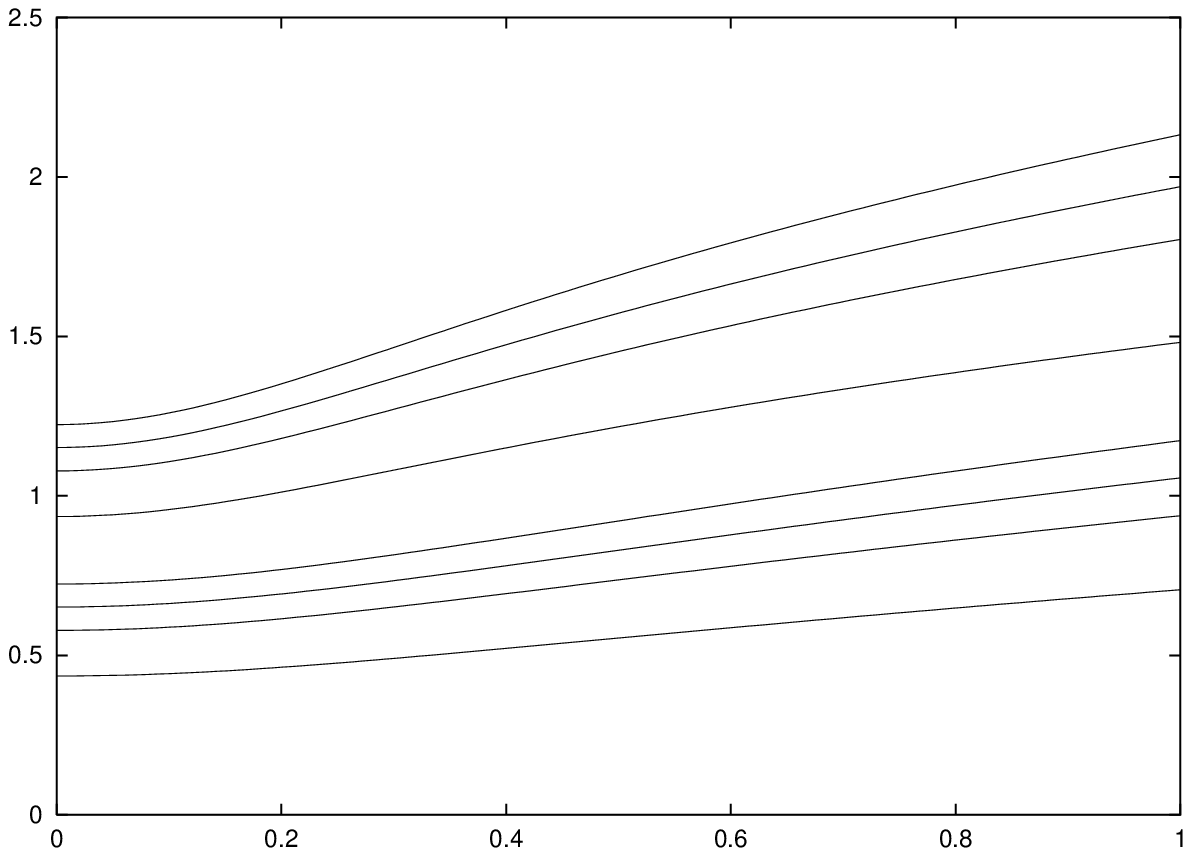} \\
(a)     &   (b)
\end{array}
$$	
$$
\begin{array}{cc}
-\sigma  &  v^2 \\
\epsfig{width=2.2in,file=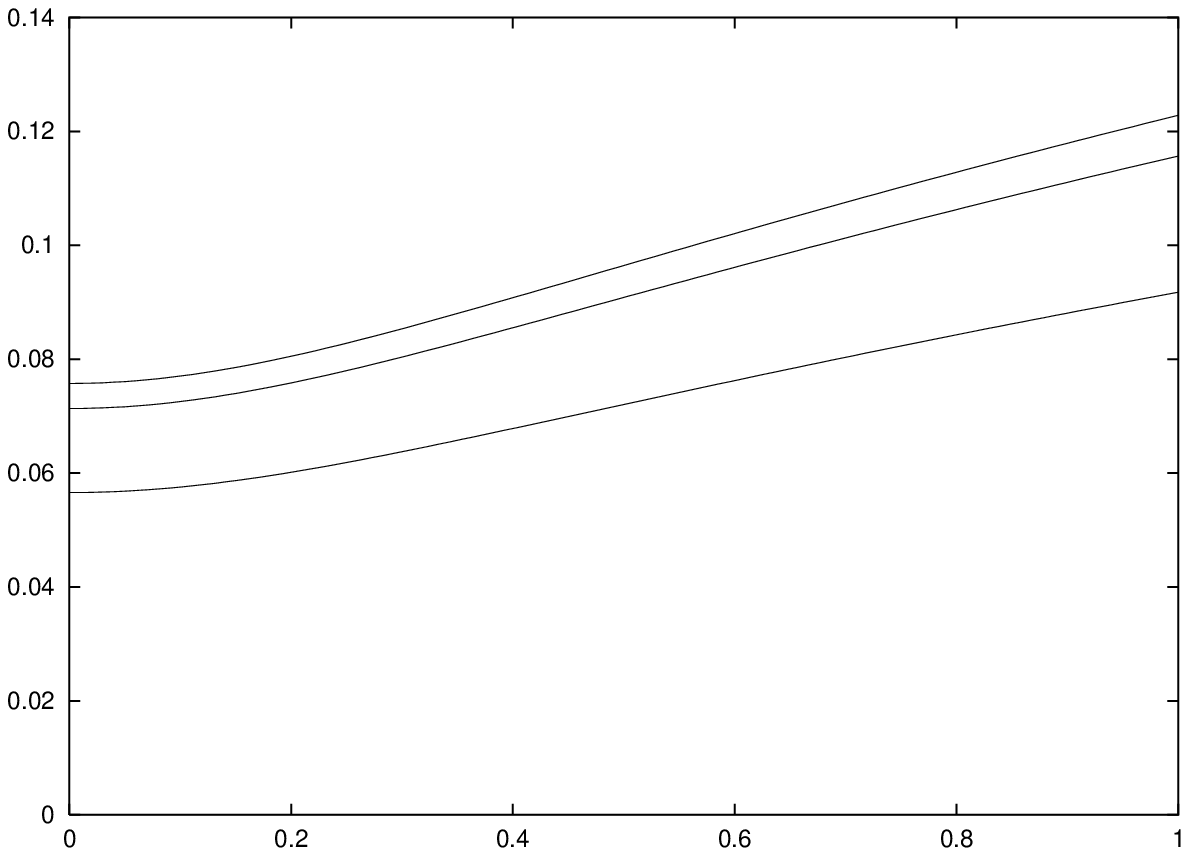} & \epsfig{width=2.2in,file=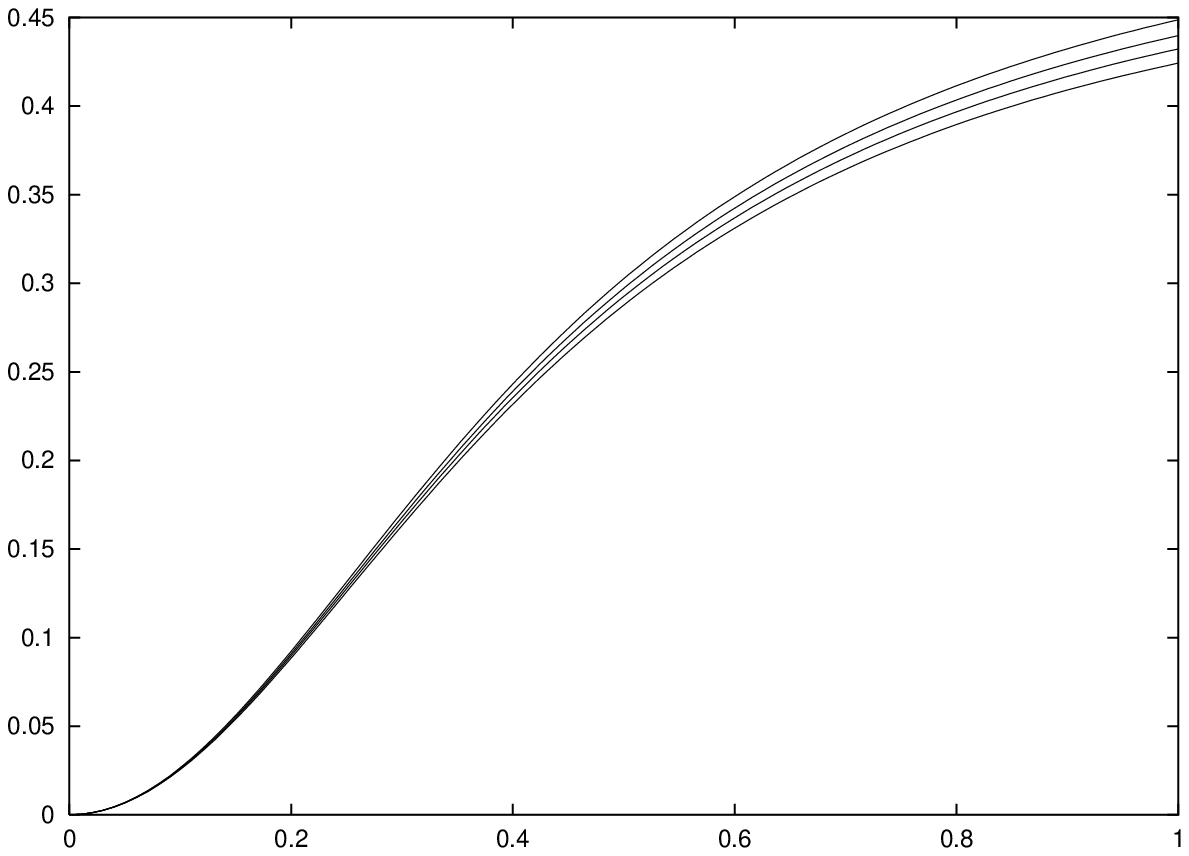} \\
(c)     &   (d)
\end{array}
$$	
$$
\begin{array}{cc}
 h^2 &  \epsilon _\pm, \ \ -\sigma _\pm \\
\epsfig{width=2.2in,file=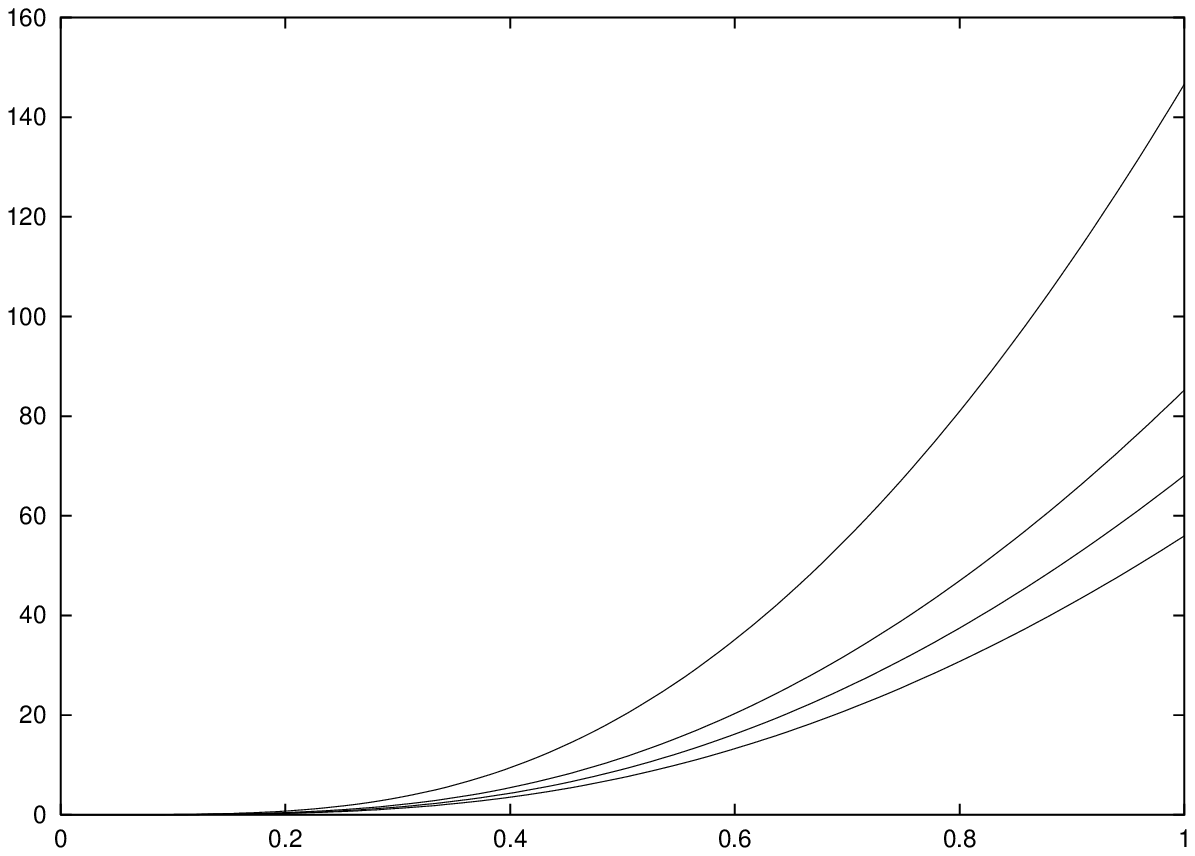} & \epsfig{width=2.2in,file=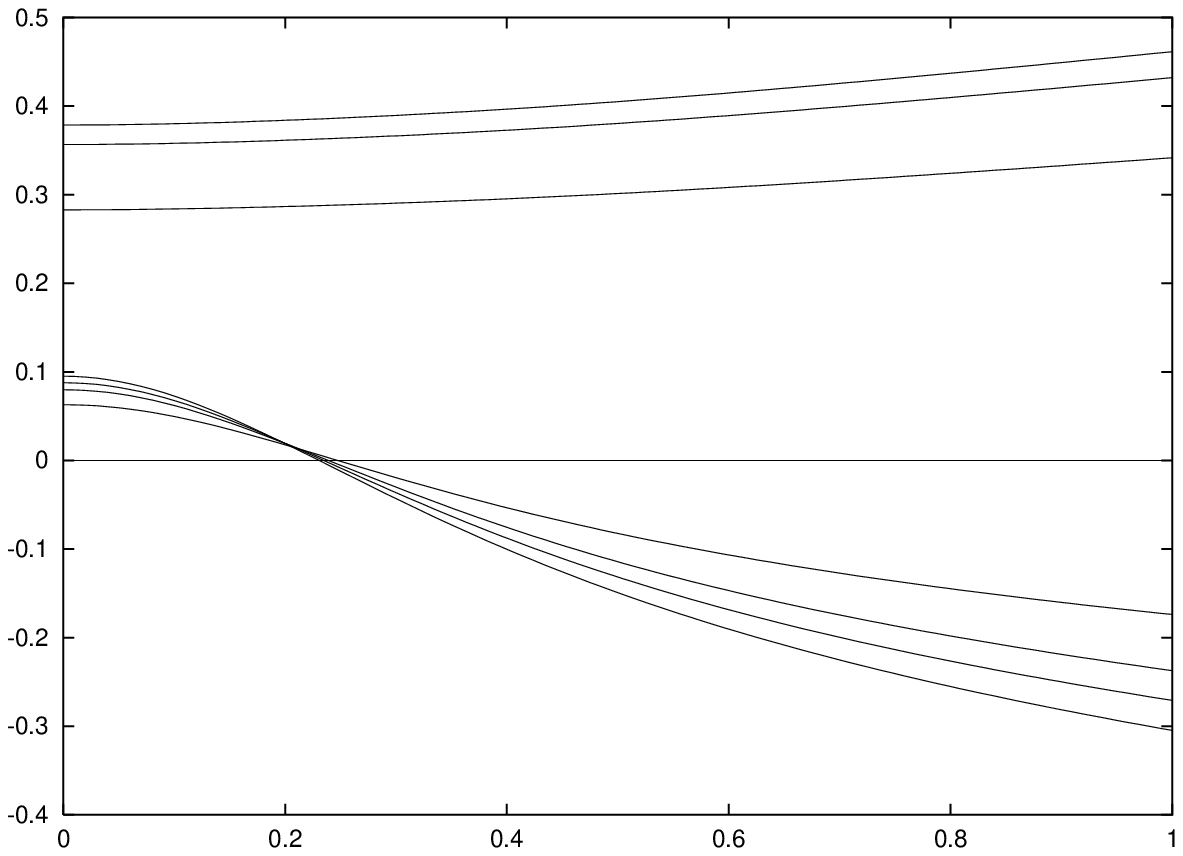}\\
(e)     &   (f)
\end{array}
$$	
\caption{For Zipoy-Voorhees-type fields we plot, as functions  of $r$, $(a)$
the energy density  $\epsilon$ and  $(b)$ the  pressures $p_\varphi$ (upper
curves moved upwards a factor of  0.5) and $p_r$  for discs  with $\alpha = 3$,
$\gamma=1.2$ and $q_1=q_2=0$ (top curves), $0.5$, $0.8$, and $1.5$  (bottom
curves), $(c)$  the electric charge density $\sigma$  for discs with $\alpha =3
$, $\gamma=1.2$ and $q_1=q_2=0$  (axis r), $0.5$, $0.8$, and $1.5$ (top curve),
$(d)$ $v^2$  for discs with $q_1=q_2=0$ (top curve), $1$, $2$, and $5$  (bottom
curve), $(e)$ the angular momentum $h^2$, and $(f)$  $\epsilon _\pm$  for
discs  with $\alpha = 3$, $\gamma=1.2$,  and $q_1=q_2=0$ (bottom curves),
$0.5$, $0.8$, and $1.5$ (top curves),  and $\sigma _\pm$ (upper curves scaled
by a factor  of 10) for $q_1=q_2=0$ (axis $r$), $0.5$, $0.8$, and $1.5$  (top
curve) and the same values de $\alpha$ and $\gamma$.}  \label{fig:zv}
\end{figure}

In Figs.  \ref{fig:zv}$(a)$ -- \ref{fig:zv}$(c)$ the plots of the quantities 
$\epsilon$, $ p_\varphi $,  $p_r$,  and  $ \sigma$  are presented for discs
with $\alpha = 3$,   $\gamma =1.2$, and  $q_1=q_2=0$, $0.5$, $0.8$, and $1.5$,
as functions of $r$. As in the previous case, we see that the energy density 
initially is a positive quantity  and then becomes negative in violation of the
weak energy condition.  Therefore  only the central region of these discs has a
physically reazonable behavior.  The another functions also have a similar 
behavior to the precedent case. We also study Zipoy-Voorhees-type discs for
other values of the parameters, but in all the cases we found a similar
behavior.

In the same way, the  relevant  quantities of the CRM are shown in the
following figures for the same values of the parameters, also as functions of
$r$. We also  consider the case when the two fluids move with equal and
opposite tangential velocities. Here the tangential velocity $v^2$ (Fig.
\ref{fig:zv}$(d)$) is always less than the light velocity.  One also finds 
that  the inclusion   of electromagnetic field makes  less  relativistic these
discs. We also obtain $ h^2$ (Fig. \ref{fig:zv}$(e)$) always  as an increasing
monotonous function of $r$ that corresponds to a stable CRM for the discs. 
Finally, in Fig. \ref{fig:zv}$(f)$ we have plotted the energy and electric
charge densities $\epsilon _\pm$ and  $\sigma _\pm$ in the particular case  
when the two streams have equal pressure or equal energy density.  We see that
$\sigma _\pm$ has a similar behavior to $\sigma$, and  $\epsilon _\pm$  to the
energy density $\epsilon$. Therefore, for  Zipoy-Voorhees-type fields we can
build  counterrotating thin disc sources only with a well-behaved central
region.

\subsection{Discs from a Bonnor-Sackfield-type solution}

The third family of solutions considered is a  Bonnor-Sackfield-type solution
which is given by
\begin{eqnarray}
e^\mu &=& \frac {2}{(a+1) e^{\gamma \cot ^{-1}u} - (a-1) e^{-\gamma \cot
^{-1}u}} , \label{eq:bs1} \\
\lambda &=&  - ( \gamma^2/2 ) \ln \left[ (u^2 + 1)/(u^2 + w^2) \right]  , \\
A_t &=& \frac{\sqrt{2} q_1 [e^{\gamma \cot ^{-1} u} - e^{-\gamma \cot ^{-1}
u}]}{(a+1) e^{\gamma \cot ^{-1} u} - (a-1) e^{-\gamma \cot ^{-1} u}}, \\
A_\varphi &=& \sqrt 2 k \gamma q_2  w, \label{eq:bs4}
\end{eqnarray}
where $a^2=1+b^2$, with  $b^2 = q_1^2 + q_2^2 $, and $\gamma$ is a real
constant. Here $q_1$ and $q_2$ are again the  electric and magnetic parameters,
respectively. $u$ and $w$ are the oblate spheroidal coordinates, related to the
Weyl coordinates by 
\begin{equation}
R^2   = k^2 (u^2+1)(1-w^2),  \quad  \quad Z  = kuw,
\end{equation}
where  $0 \leq u \leq \infty$ and $0  \leq w \leq 1 $. In this case we must
choose  $0 < \alpha < 1$ and the shell would be  located at $u =  \alpha / k >
0$ and $w= \sqrt {1-r^2}$, with $k= \sqrt{1-\alpha ^2 }$. As in the precedent
cases this solution can  be generated, in these coordinates, following  Ernst's
method \cite{E2} from the   Bonnor-Sackfield  vacuum solution \cite{BS}.

From the above expressions we can compute the physical quantities associated
with the discs. We obtain
\begin{eqnarray}
\epsilon &=& \varrho - p_r -p_ \varphi,   \\
p_r  & =& p_0 [1 - k^2 r^2] ^{-(\gamma ^2+1)/2} \\
p _\varphi & = & \left [ \frac {1+ k^2\gamma^2 r^2}{1- k^2r^2} \right] p_r, \\
\sigma &=&  -  \frac { \gamma k q_1  p_0}{\sqrt 2 \alpha ^2} \ p_r, 
\end{eqnarray} 
where
\begin{equation}
p_0 =  \frac {4 \alpha}{(a+1)e^{\gamma \cot^{-1} u }-(a-1)e^{-\gamma \cot^{-1}
u }}
\end{equation}
is a constant and 
\begin{equation}
\varrho =   \frac {\gamma k p_0}{2 \alpha ^2} \left[ (a+1)e^{\gamma \cot^{-1} u
}+(a-1)e^{-\gamma \cot^{-1} u }\right ]  p_r.
\end{equation}
Here  we also must choose $\gamma >0$ so that $\varrho \geq 0$  and the
solution satisfies the strong energy condition.   $p_0$ is also  a positive
constant for these values of $\gamma$. 

\begin{figure}
$$
\begin{array}{cc}
\epsilon  &  p_\varphi, \ \ p_r \\
\epsfig{width=2.2in,file=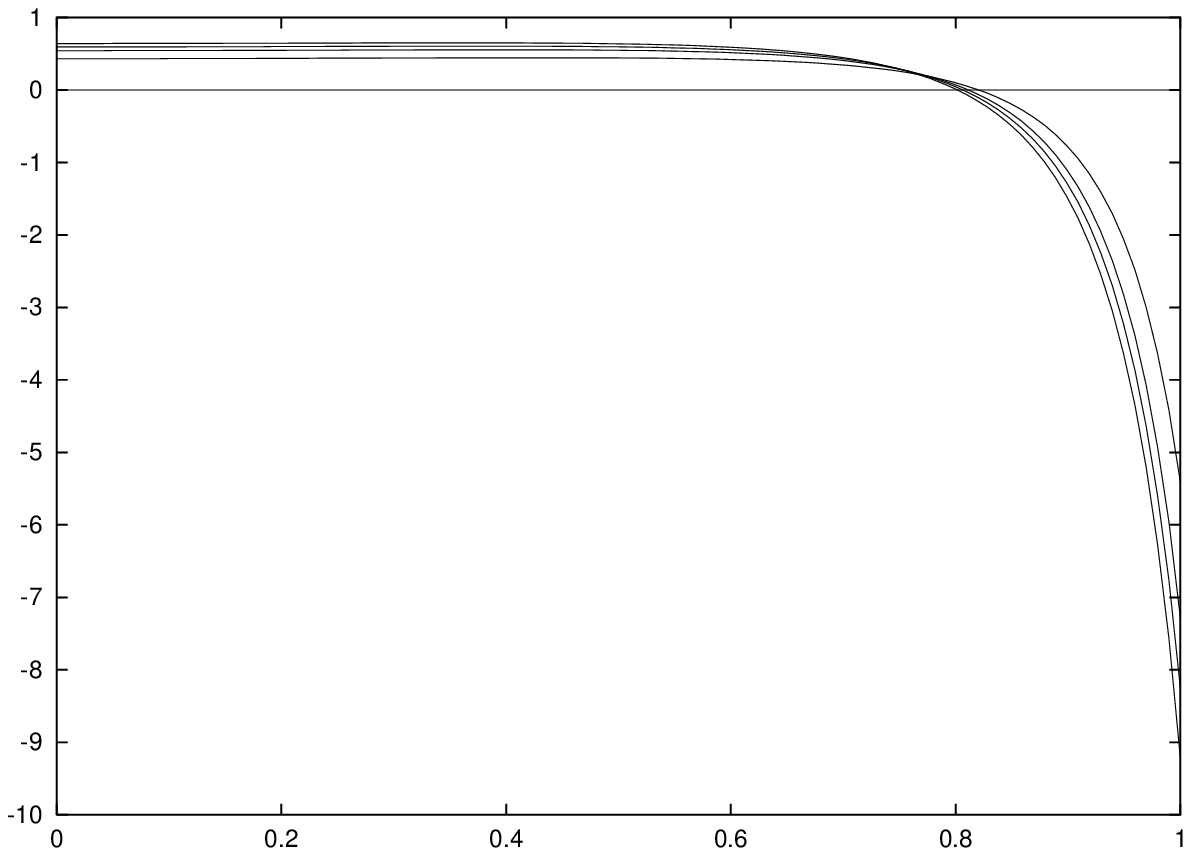} & \epsfig{width=2.2in,file=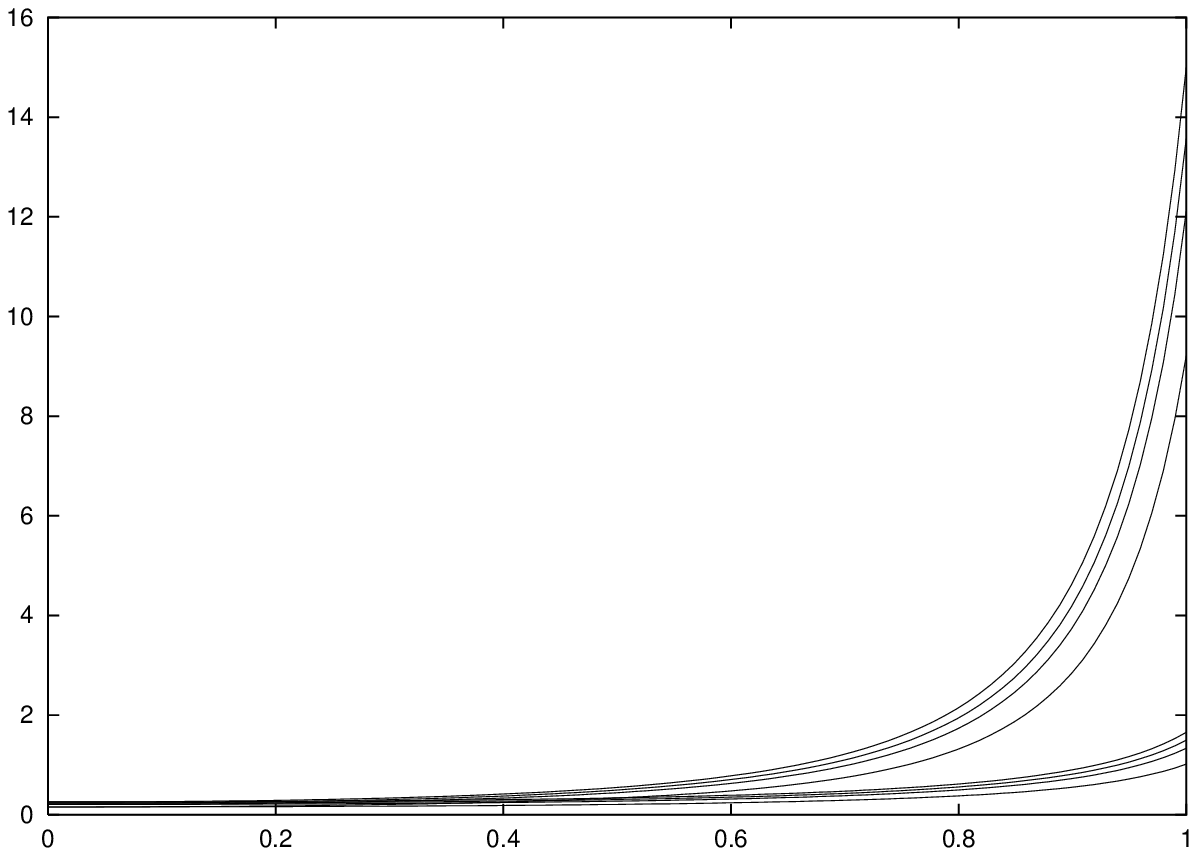} \\
(a)     &   (b)
\end{array}
$$	
$$
\begin{array}{cc}
-\sigma &  v^2 \\
\epsfig{width=2.2in,file=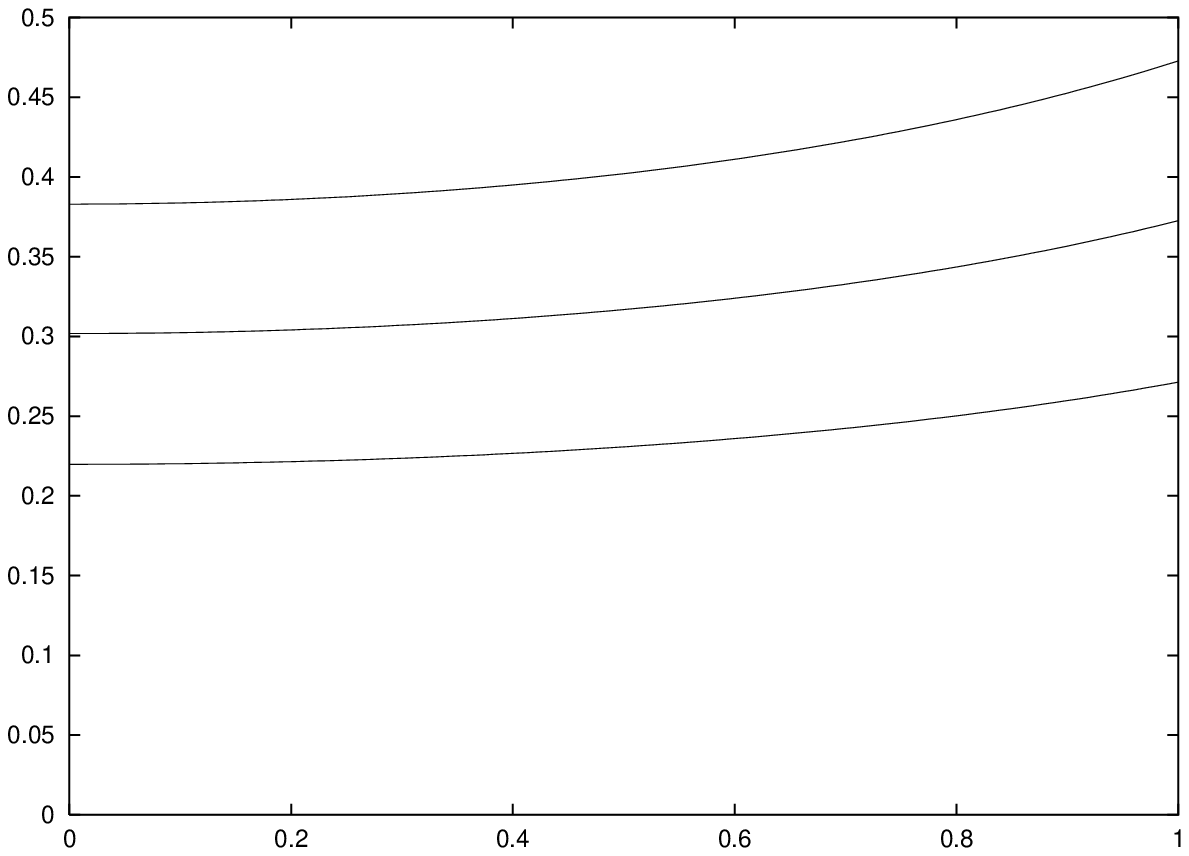} & \epsfig{width=2.2in,file=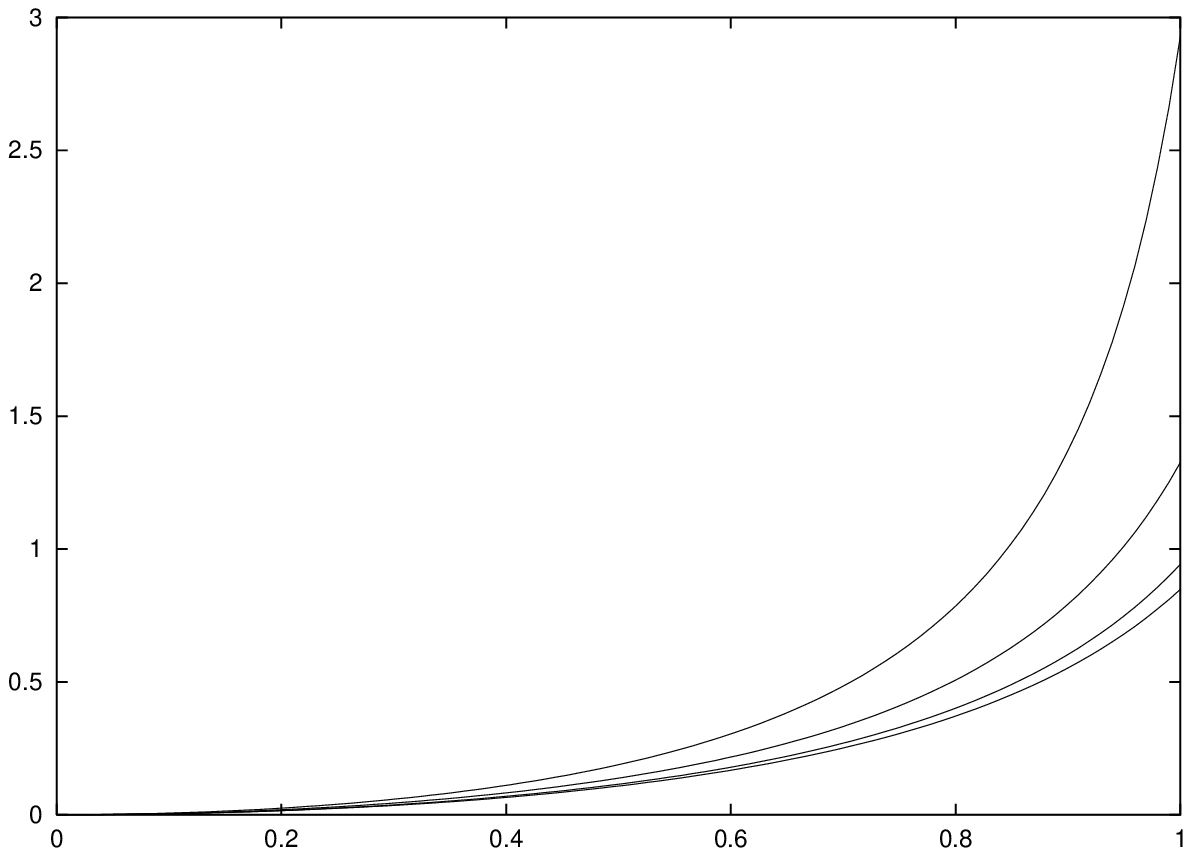} \\
(c)     &   (d)
\end{array}
$$	
$$
\begin{array}{cc}
 h^2 &  \epsilon _\pm, \ \ -\sigma _\pm \\
\epsfig{width=2.2in,file=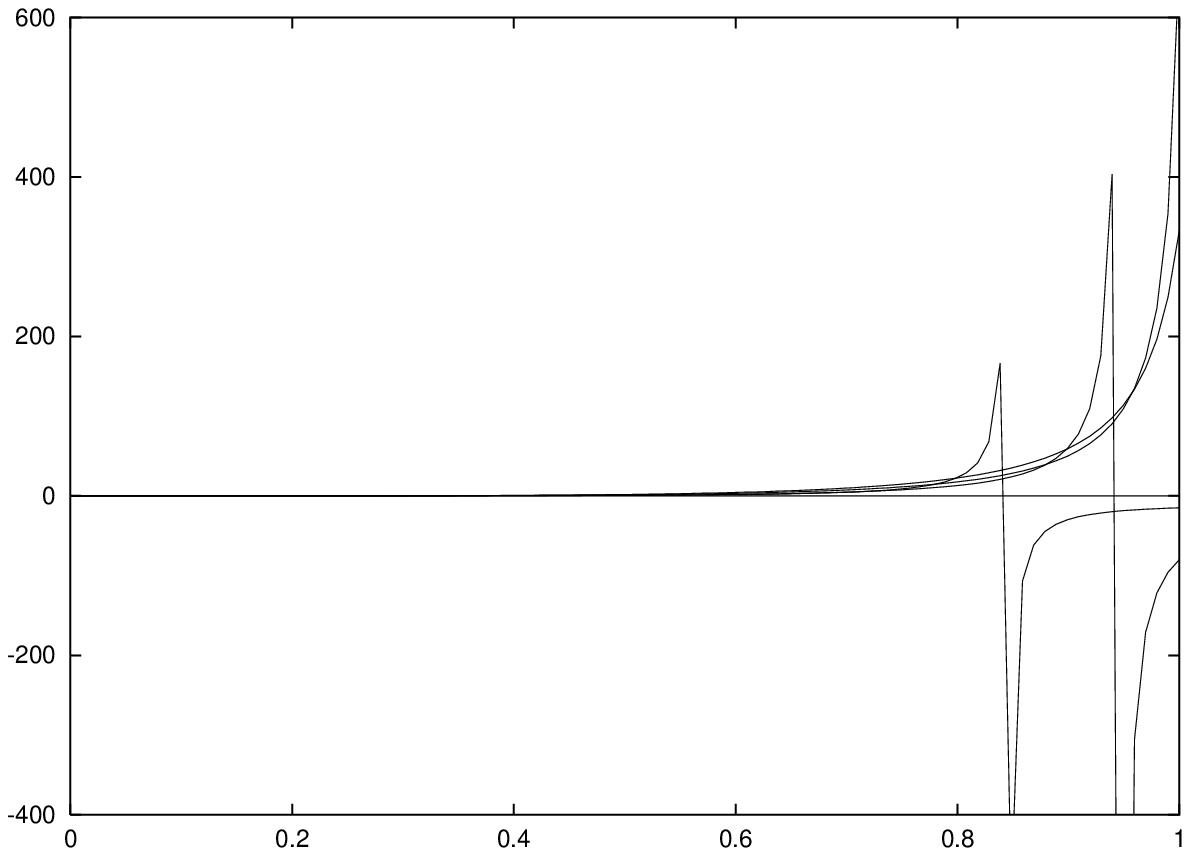} & \epsfig{width=2.2in,file=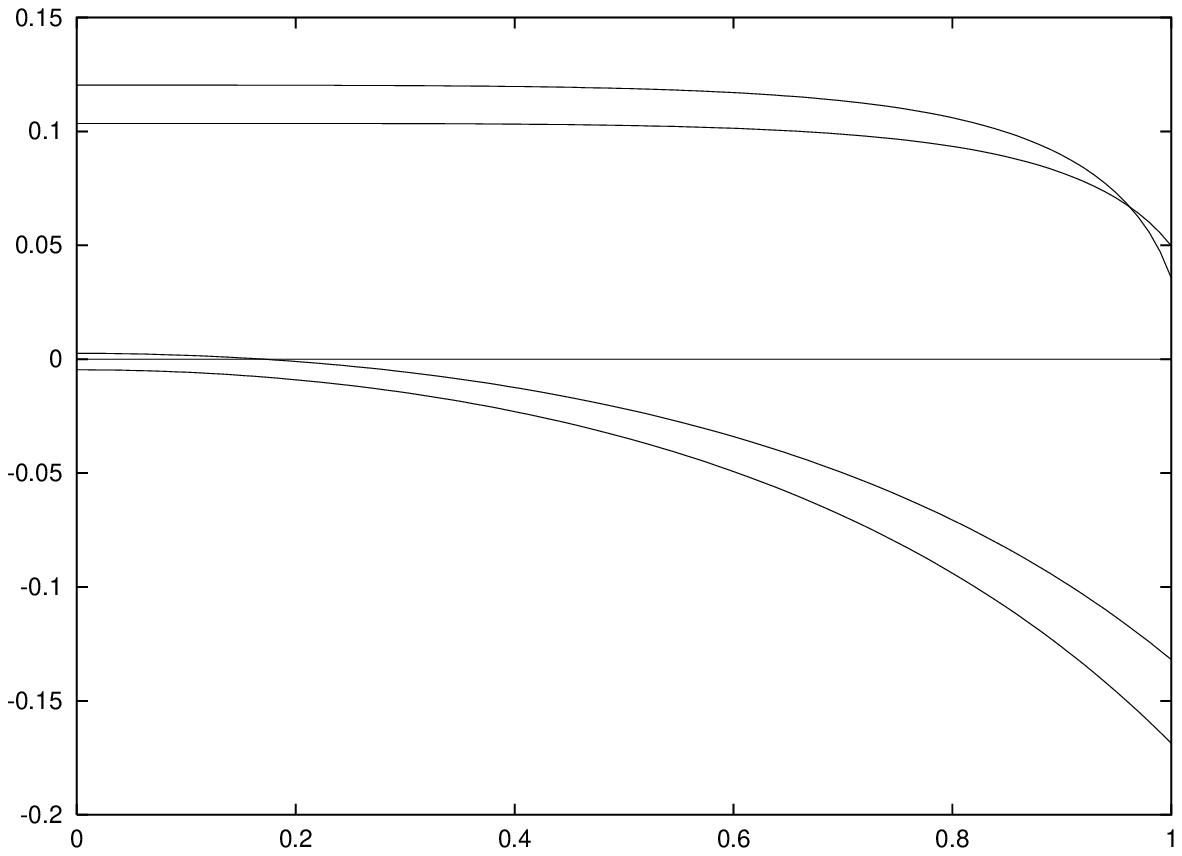}\\
(e)     &   (f)
\end{array}
$$	
\caption{For Bonnor-Sackfield-type fields we plot,  as functions of $r$, $(a)$
the energy density  $\epsilon$ and $(b)$ the  pressures $p_\varphi$ (upper
curves)  and $p_r$  for discs  with $\alpha = 0.5$, $\gamma=1.3$ and
$q_1=q_2=0$ (top  curves), $0.5$, $0.8$, and $1.5$ (bottom curves), $(c)$  the 
electric charge  density  $\sigma$ for discs  with $\alpha = 0.5 $,
$\gamma=1.3$ and $q_1=q_2=0$ (axis r),  $0.5$, $0.8$,  and $1.5$ (top curve),
$(d)$ $v^2$ for discs  with $\alpha = 0.9$, $\gamma=1$ and $q_1=q_2=3$ (bottom
curve), $5$, $8$, and $10$ (top curve), $(e)$ the angular momentum  $h^2$  for
discs with $\alpha = 0.9$,  $\gamma=1$, and  $q_1=q_2=3$, $5$ (sharp curves),
$8$, and $10$ (bottom curve), and $(f)$  $\epsilon _\pm$ (lower curves) and
$\sigma _\pm$ for  discs   with $\alpha = 0.9$,  $\gamma=1$, and  $q_1=q_2=8$
and $10$  (top and bottom curves, respectively).} \label{fig:bs}
\end{figure}

In Figs.  \ref{fig:bs}$(a)$ -- \ref{fig:bs}$(c)$ the plots of the above
quantities are presented for discs with $\alpha = 0.5$,   $\gamma =1.3$, and 
$q_1=q_2=0$, $0.5$, $0.8$, and $1.5$, as functions of $r$. As in the previous
cases, we see that the energy density  initially is a positive quantity  and
then becomes negative in violation of the weak energy condition.  Therefore 
only the central region of these discs has a physically reazonable behavior. 
The another functions also have a similar  behavior to the earlier cases. We
also study Bonnor-Sackfield-type discs for other values of the parameters, but
in all the cases we found a similar behavior.

Again, the  relevant  quantities of the CRM are shown in the following figures,
also as functions of $r$. We also  consider the case when the two fluids move
with equal and opposite tangential velocities. Here the tangential velocity
$v^2$ (Fig. \ref{fig:bs}$(d)$) is not always less than the light velocity.
However the inclusion of  electromagnetic field can make the velocities of the
particles smaller than the light velocity.  In some cases $ h^2$  (Fig.
\ref{fig:bs}$(e)$) presents strong changes in the  slope at certain value of
$r$, which means that there is a strong instability there. For other values of
the parameters $ h^2$ is an increasing monotonous function of $r$ that
corresponds to a stable CRM for the discs.

Finally, in Fig. \ref{fig:bs}$(f)$ we have plotted the energy and electric
charge densities $\epsilon _\pm$ and  $\sigma _\pm$ in the particular case  
when the two streams have equal pressure or equal energy density. We see that  
$\epsilon _\pm$  has a similar behavior to the energy density $\epsilon$.
Therefore, for  Bonnor-Sackfield-type fields we can build  counterrotating thin
disc sources only with a well-behaved central region.

\subsection{Discs from a charged and magnetized Darmois  solution}

Finally, other family of solutions to the Einstein-Maxwell  equations is a
charged and magnetized Darmois solution, which is given by
\begin{eqnarray}
e ^\mu  &=& \frac{a^2 x^2 - b^2 y^2 - 1}{(a x + 1)^2 - b^2 y^2},
\label{eq:kerr1} \\
\lambda &=& 2 \ln \left[\frac{a^2 x^2 - b^2 y^2 - 1}{a^2 (x^2 - y^2)} \right],
\\
A_t &=& \frac {2 \sqrt 2 q_1 y}{(ax + 1)^2 - b^2 y^2},  \\
A_\varphi  &=& -\frac{\sqrt 2 k q_2 (1 - y^2)(ax + 1)}{a(a^2 x^2 - b^2 y^2 -
1)}, \label{eq:kerr4}
\end{eqnarray}
where $a^2=1+b^2$, with $b^2 = q_1^2 + q_2^2 $. Here $q_1$ and $q_2$ are also
the  electric and magnetic parameters, respectively. $x$ and $y$ are  the
prolate spheroidal coordinates, given by  (\ref{eq:coorp}). Again we must
choose  $\alpha >1$ and the shell would be located at $x =  \alpha / k > 1$ and
$y= \sqrt {1-r^2}$, with  $k= \sqrt {\alpha ^2 -1}$. For $b=0$ this solution 
also  reduces to the Darmois metric. Moreover, for  $q_2=0$ we have a Kerr-type
electrostatic solution. This  solution can   be generated, in these
coordinates,  using a well-known theorem  proposed by Bonnor  (see Ref.
\cite{B2}) from  the Kerr  vacuum solution. And for $q_1=0$  one  obtains its 
magnetostatic equivalent.  This  solution is asymptotically flat and  was
firts  studied by Bonnor \cite{B3}  in a different  system of coordinates  and 
describes the field of a massive magnetic dipole.

The physical quantities associated with the discs  can now be written as 
\begin{eqnarray}
\epsilon &=& \varrho - p_r -p_ \varphi,   \\
p_ \varphi &=& \left[ \frac {a^2 + k^2(4+b^2)r^2}{(1+k^2 r^2)(a^2 + k^2b^2r^2)}
\right] p_r,   \\
p_r &=& \frac {2 \alpha a^4 (1+k^2 r^2)^{3/2}}{(a^2 + k^2b^2r^2)[(\alpha a +
k)^2 - k^2b^2(1-r^2)]},  \\
\sigma &=& - \left[  \frac {4 \sqrt 2 a q_1 (\alpha a + k) (1-r^2)^{1/2} }
{\alpha (a^2 + k^2b^2r^2) [(\alpha a + k)^2 - k^2b^2(1-r^2)] } \right]  p_r,  
\\
\mbox{\sl j} &=& - \left[  \frac {\sqrt 2 q_2 k r^2}{4a} \right] \varrho,
\end{eqnarray}
where
\begin{equation}
\varrho = \frac{4a k }{\alpha (a^2 + k^2b^2r^2) } \left[ \frac {(\alpha a +
k)^2 + k^2b^2(1-r^2)}{(\alpha a + k)^2 - k^2b^2(1-r^2)} \right] p_r.
\end{equation}

\begin{figure}
$$
\begin{array}{cc}
\epsilon  &  p_\varphi, \ \ p_r \\
\epsfig{width=2.2in,file=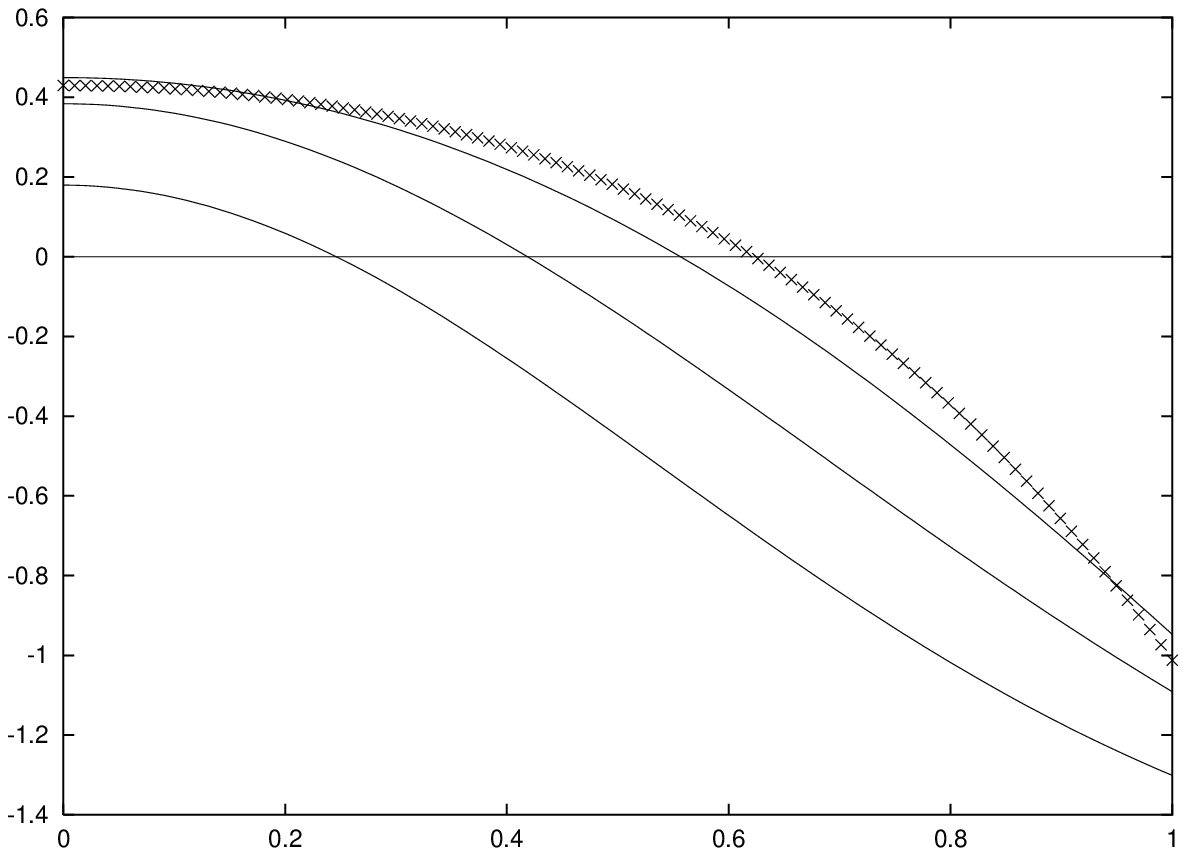} & \epsfig{width=2.2in,file=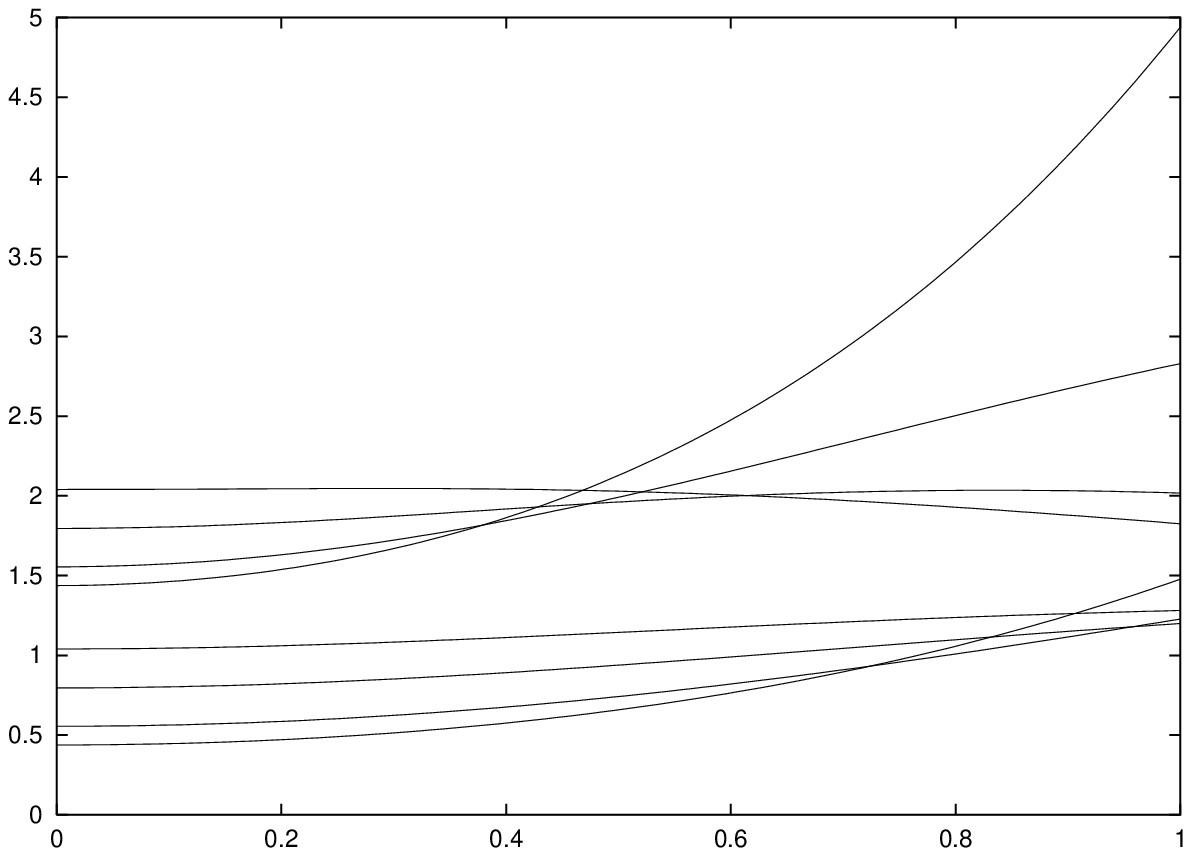} \\
(a)     &   (b)
\end{array}
$$	
$$
\begin{array}{cc}
-\sigma, \  -\mbox{\sl j} &  v^2 \\
\epsfig{width=2.2in,file=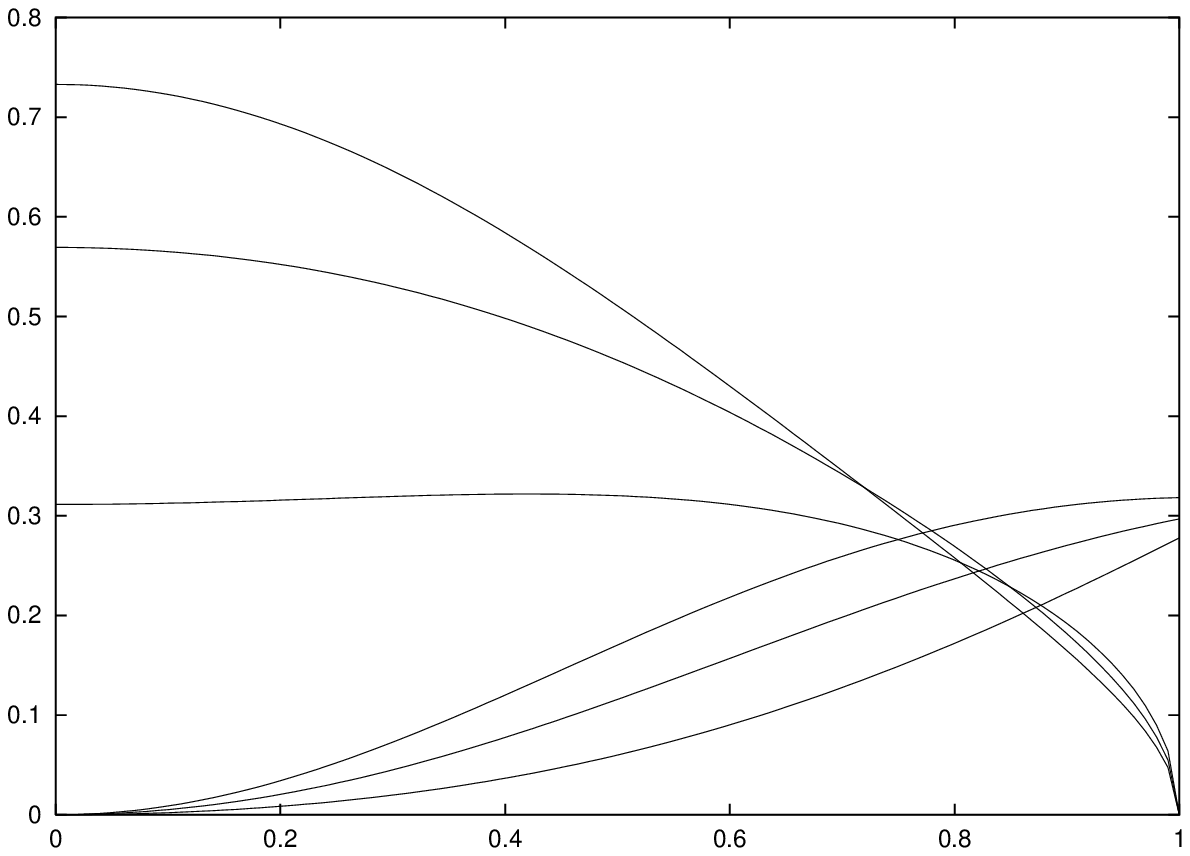} & \epsfig{width=2.2in,file=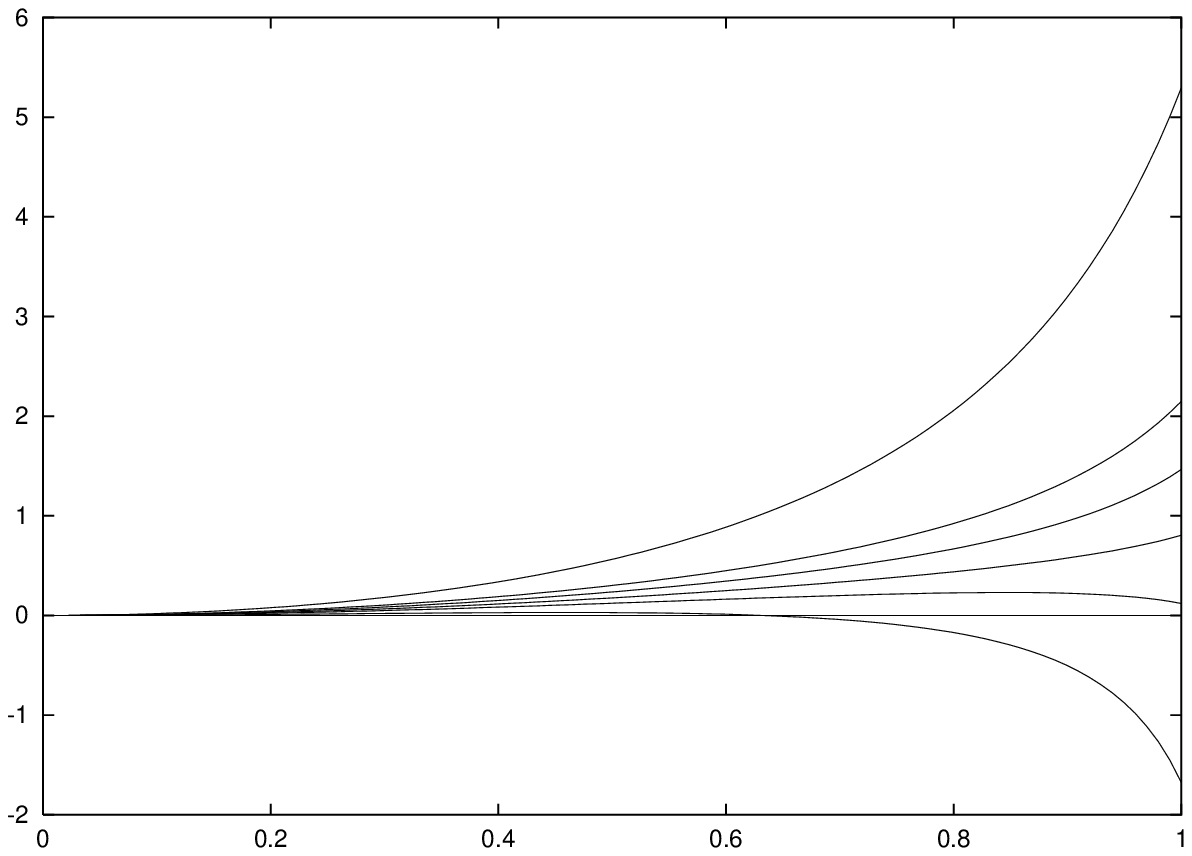} \\
(c)     &   (d)
\end{array}
$$	
$$
\begin{array}{cc}
 h^2 &  \epsilon _\pm, \ \ -\sigma _+, \  \sigma _-  \\
\epsfig{width=2.2in,file=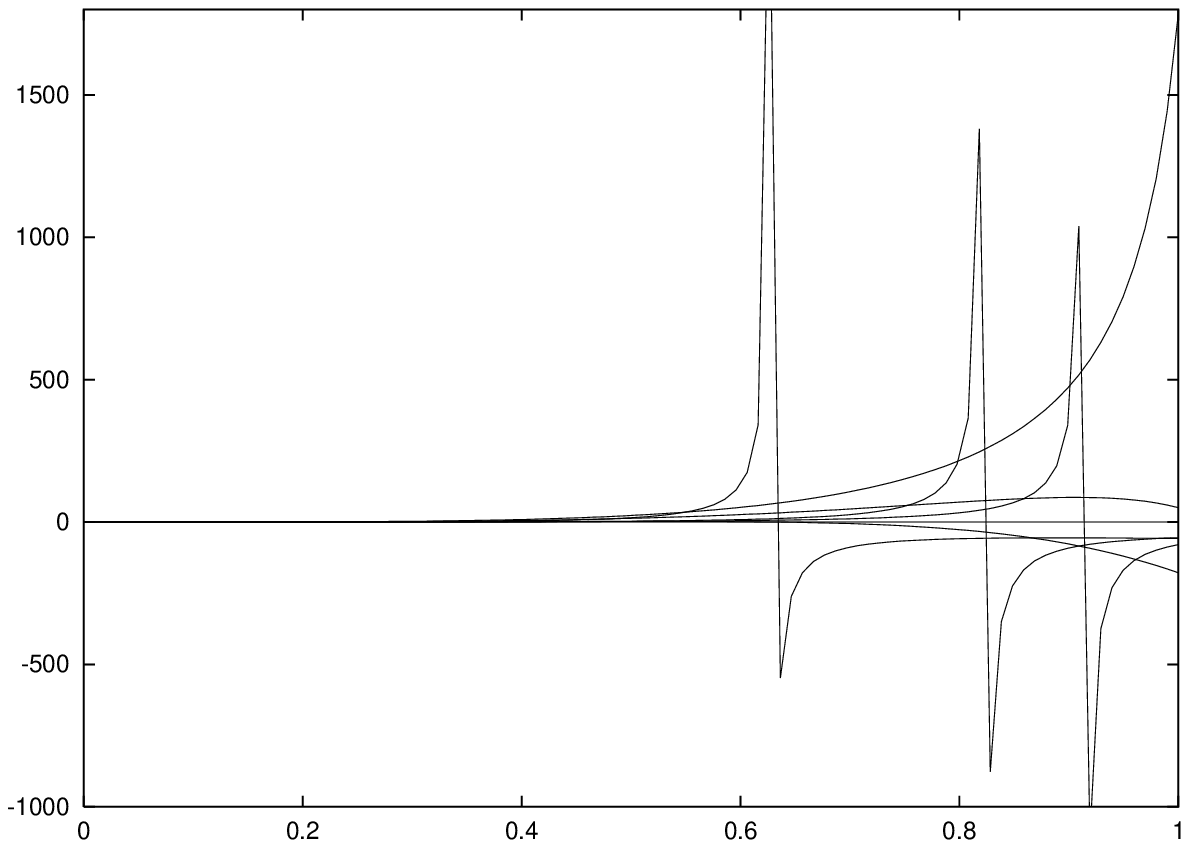} & \epsfig{width=2.2in,file=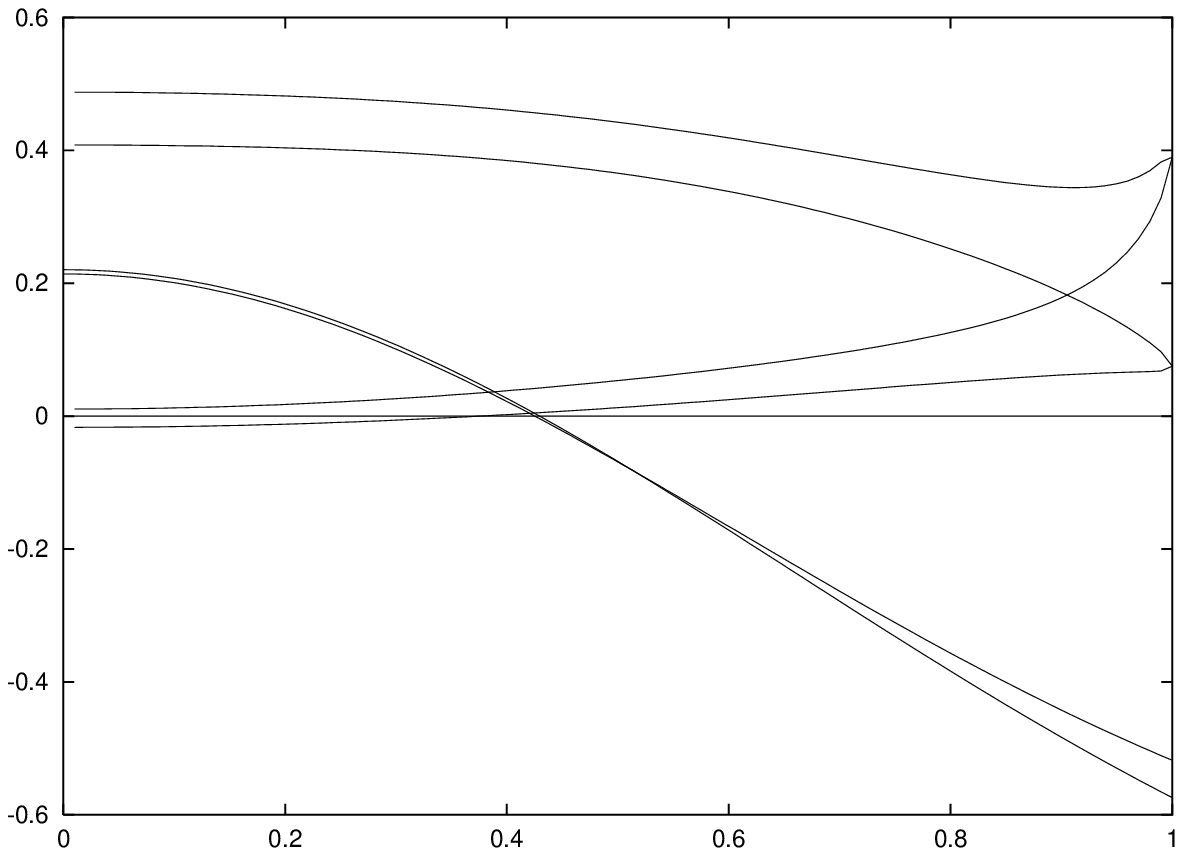}\\
(e)     &   (f)
\end{array}
$$	
\caption{For charged and magnetized Darmois fields we plot, as functions of
$r$, $(a)$ the energy density  $\epsilon$ for  discs with $\alpha = 1.5$ and
$q_1=q_2=0$ (curve with  crosses), $0.5$, $1.0$, and $1.5$ (bottom curve),
$(b)$ the  pressures $p_\varphi$ (upper curves) and $p_r$ for discs with
$\alpha = 1.5$ and $q_1=q_2=0$, $0.5$, $1.0$,  and $1.5$ (top curves at the
border of the discs), $(c)$   $\sigma$ (upper curves) and $\mbox{\sl j}$ for 
discs with $\alpha =1.5$ and $q_1=q_2=0$  (axis r), $0.5$, $1.0$,  and $1.5$
(top curves), $(b)$ $v^2$ for discs  with  $\alpha =1.5$ and $q_1=q_2=0$ (top
curve), $0.5$, $0.6$,  $0.7$, $0.8$, and $1.0$ (bottom curve), $(e)$ the
angular  momentum $h^2$ for discs with $\alpha =1.5$ and $q_1=q_2=0$,  $0.5$,
$0.6$ (sharp curves), $0.7$ (top curve), $0.8$, and  $1.0$ (bottom curve),
where the  last curves have been scaled  by a factor of 20, and $(f)$ $\epsilon
_\pm$, $\sigma _+$  (upper curves), and $\sigma _- $ (lower curves) for discs
with $\alpha =1.5$ and $q_1=q_2=0.7$ and  $0.8$ (top, bottom and  bottom
curves, respectively).} \label{fig:k}
\end{figure}

In Figs.  \ref{fig:k}$(a)$ -- \ref{fig:k}$(c)$ the plots of the above 
quantities  are shown for discs with $\alpha = 1.5$ and  $q_1=q_2=0$, $0.5$,
$1.0$, and $1.5$, as functions of $r$.  We see that these discs also have a
central region well behaved  which satisfies the weak and strong energy
conditions and border region where $\epsilon < 0$, in violation of the weak
energy condition. We also computed these functions for other values of the
parameters, but in all the  cases we found a similar behavior.

Equally, the  relevant  quantities of the CRM are shown in the following
figures, also as functions of $r$. We also  consider the case when the two
fluids move with equal and opposite tangential velocities. We  see that the
inclusion of  electromagnetic field can make the velocities of the particles
smaller than the light velocity. Here $v^2$  [Fig. \ref{fig:k}$(d)$] can also
take  negative values. In some cases $ h^2$  [Fig. \ref{fig:k}$(e)$)] presents
strong changes in the  slope at certain values of $r$, which means that we have
a strong instability there, and  also we find  regions with negative slope,
showing that the CRM can not be applied for these values of the parameters. 
However for $q_1=q_2=0.7$  $ h^2$ is always  an increasing monotonous function
of $r$ that corresponds to a stable CRM for this disc. 

Finally, in Fig. \ref{fig:k}$(f)$ we have plotted  $\epsilon _\pm$,  $\sigma
_+$, and  $\sigma _-$ in the particular case   when the two streams have equal
pressure or equal energy density, for values of the parameters for which
$v^2<1$. We see that $\epsilon _\pm$   has a similar behavior  to the energy
density $\epsilon$. Therefore, for this solution we can build  counterrotating 
thin disc sources only with a well-behaved central region. 

\section{Discussion}\label{sec:discu}

A detailed study was presented of the counterrotating model for generic  
electrovacuum static  axially symmetric relativistic thin discs with nonzero
radial pressure. A general constraint over the counterrotating tangential
velocities was found, needed to cast the surface energy-momentum tensor of the
disc in such a way that it can be interpreted as the superposition of two
counterrotating  charged perfect fluids. The constraint found is completely
equivalent to the necessary and sufficient condition obtained in Ref.
\cite{FMP}.  We show that, in general, it is not possible to take the two
counterrotating fluids as circulating along electrogeodesics neither take the
two counterrotating tangential velocities as equal and opposite. We also have
obtained explicit expressions for the energy densities,  current densities, and
velocities   of the counterrotating streams in terms of the energy density,
azimuthal pressure, and planar current density of the discs, that are also
equivalent to the correspondig expressions in Ref. \cite{FMP}.

Four families of models of counterrotating charged discs were considered in the
present work based on simple solutions to the vacumm Einsteins-Maxwell
equations  in the static axisymmetric case generated by conventional
solution-generating tecniques \cite{KSHM}. We  saw  that the presence  of an
electromagnetic field can make the counterrotating tangential  velocities of
the particles smaller than the light velocity, and also can stabilizes againts
radial perturbations  the CRM.  The discs constructed   have  a well-behaved
central region   which satisfies the weak and strong energy conditions and
border region where the energy density is negative, in violation of the weak
energy condition. 

Finally, the generalization of the counterrotating model presented here to the
case  of  rotating thin discs with or without radial pressure  in presence of
electromagnetic fields is being considered.

\ack

The authors want to thank the finantial support from  COLCIENCIAS, Colombia.
G.G.R. also wants to thank the the  Centro Nacional de C\'alculo Cient\'\i
fico, Universidad de  los Andes, Venezuela, for finantial support, and the warm
hospitality of the Centro de Astrof\'\i sica Te\'orica,  Universidad de los
Andes, Venezuela, where part of this work was performed.


\end{document}